\documentclass[pre,reprint,twocolumn,footinbib,floatfix]{revtex4-2}

\usepackage[T1]{fontenc}
\usepackage[utf8]{inputenc}

\usepackage{amsmath}
\usepackage{amsfonts}
\usepackage{amssymb}
\usepackage{mathtools}
\usepackage{empheq}
\usepackage{mathrsfs} 
\usepackage{titlesec}
\usepackage{indentfirst}

\usepackage{physics}
\let\exp\exponential
\let\ln\naturallogarithm

\allowdisplaybreaks

\usepackage{graphicx}
\usepackage[usenames,dvipsnames]{xcolor}

\newenvironment{red}{\begingroup\color{red!70!black}}{\endgroup}

\definecolor{pdarkblue}{rgb}{0.1797, 0.1875, 0.5703}
\usepackage{hyperref}
\hypersetup{
    colorlinks=true,
    citecolor=pdarkblue,
    linkcolor=pdarkblue,
    urlcolor=pdarkblue,
}
\usepackage[all]{hypcap}
\usepackage[varg]{txfonts}
\usepackage{newtxtext}

\AtBeginDocument{
\fontsize{10.5pt}{12.36pt}\selectfont
\fontdimen2\font=2.9pt 
}
\usepackage{titlesec}
\titleformat*{\section}{\centering\fontsize{10.5pt}{\baselineskip}\selectfont\bfseries}
\titleformat*{\subsection}{\centering\fontsize{10.5pt}{\baselineskip}\selectfont\bfseries}
\titleformat*{\subsubsection}{\centering\fontsize{10.5pt}{\baselineskip}\selectfont\itshape}
\titlespacing*{\section}{\linewidth}{2.5em}{0.4em}
\titlespacing*{\subsection}{\linewidth}{1.5em}{0.4em}
\titlespacing*{\subsubsection}{\linewidth}{1.5em}{0.4em}

\newcommand{\kB}{k_{\text{\normalfont\scshape b}}}
\newcommand{\DKL}{D_{\text{\normalfont\scshape kl}}}
\newcommand{\eq}{\mathrm{eq}}
\newcommand{\st}{\mathrm{st}}
\newcommand{\epr}{{\tilde{\sigma}}}
\newcommand{\pns}{{\tilde{p}}}
\newcommand{\bmetans}{{\tilde{\bm\eta}}} 
\renewcommand{\H}{{\mathcal{H}}}
\newcommand{\Hm}{{\mathcal{H}_{\mathrm m}}}
\newcommand{\F}{{\mathcal{F}}}
\newcommand{\inter}{\mathrm{int}}
\newcommand{\pTE}{p_{\scriptstyle\text{\normalfont\scshape te}}}
\newcommand{\gen}{\mathscr{L}}
\newcommand{\trp}{\mathsf{T}}
\newcommand{\llangle}{\langle\!\langle}
\newcommand{\rrangle}{\rangle\!\rangle}
\def\SetMake#1|#2?{\left\{#1\,\middle|\,#2\right\}}

\newcommand{\sca}{\beta}
\newcommand{\scb}{\gamma}
\newcommand{\scc}{\mu}
\newcommand{\scd}{\nu}

\usepackage{bm}

\begin{document}

\title{Inferring nonequilibrium thermodynamics from tilted equilibrium\texorpdfstring{\\[0.5pt]}{ } using information-geometric Legendre transform}
\date{\today}

\author{Naruo Ohga}
\email{naruo.ohga@ubi.s.u-tokyo.ac.jp}
\author{Sosuke Ito}
\affiliation{Department of Physics, Graduate School of Science, The University of Tokyo, 7-3-1 Hongo, Bunkyo-ku, Tokyo 113-0033, Japan}

\begin{abstract}
Nonstationary thermodynamic quantities depend on the full details of nonstationary probability distributions, making them difficult to measure directly in experiments and numerics. 
We propose a method to infer thermodynamic quantities in relaxation processes by measuring only a few observables, using additional information obtained from measurements in tilted equilibrium, i.e., equilibrium with external fields applied. 
Our method is applicable to arbitrary classical stochastic systems, possibly underdamped, that relax to equilibrium.
The method allows us to compute the exact value of the minimum entropy production (EP) compatible with the nonstationary observations, giving a tight lower bound on the true EP\@. 
Under a certain additional condition, it also allows the inference of the EP rate, thermodynamic forces, and a constraint on relaxation paths.
Our method uses a Legendre transform of EP at the level of probability distributions, which we develop based on a similar Legendre transform in information geometry. 
\end{abstract}

\maketitle

\section{Introduction}

Relaxation processes are ubiquitous in nature, and they undergo various nonstationary probability distributions before relaxing to the final stationary distribution. For example, biological systems such as biochemical signaling pathways~\cite{Bialek2012Biophysics,Bhalla1999EmergentProperties} and neurons~\cite{dayan2001TheoreticalNeuroscience,Wark2007SensoryAdaptation} respond to external signals in a transient manner to convey information. Relaxation processes also include various nontrivial physical phenomena, such as nonmonotonic relaxations~\cite{polettini2013nonconvexity,lu2017nonequilibrium,Kumar2020ExponentiallyFaster}, slow glassy relaxations~\cite{Berthier2011TheoreticalPerspectiveOnTheGlass}, and sudden cutoff relaxations~\cite{Diaconis1996Cutoff,Kastoryano2012Cutoff}.

These processes are out of equilibrium and thus have inevitable thermodynamic costs. According to stochastic thermodynamics~\cite{schnakenberg1976network,seifert2012stochastic,sekimoto2010stochastic}, thermodynamic costs such as entropy production (EP) and thermodynamic forces depend on the full details of nonstationary probability distributions. Therefore, to obtain thermodynamic quantities directly in experiments and numerical simulations, we need to measure all the details of probability distributions by accumulating many realizations of the same relaxation process and computing the histogram at each time point. This nonstationary measurement is practically impossible for systems with more than a few states.

To reduce the demand for nonstationary measurements, we need to develop indirect methods to infer thermodynamic quantities from measurable partial information. Such methods have been actively studied in stochastic thermodynamics, collectively referred to as \textit{thermodynamic inference}~\cite{seifert2019stochastic}. Previous studies have used various types of measurable partial information, such as 
raw time series data~\cite{roldan2010estimating,Roldan2012EntropyProduction,muy2013noninvasive}, 
correlation and response functions~\cite{Harada2005PRL,Harada2006PRE}, 
coarse-grained states~\cite{rahav2007fluctuation,esposito2012stochastic,Bo2014EntropyProduction,Skinner2021Improved}, 
partial transitions~\cite{shiraishi2015fluctuation,polettini2017effective,bisker2017hierarchical}, 
current observables~\cite{gingrich2017inferring,dechant2021improving}, 
mean local velocities~\cite{lander2012noninvasive,frishman2020learning,Gnesotto2020Learning},
arbitrarily weighted local currents~\cite{Li2019Quantifying,manikandan2020inferring,otsubo2020estimating,vu2020entropy,Manikandan2021Quantitative,Kim2020Learning,Kim2022Estimating,Otsubo2022estimating,Lee2023Multidimensional}, 
sojourn probabilities of tubular paths~\cite{Kappler2022Measurement}, 
the Fisher information~\cite{ito2020prx}, 
and waiting time distributions of transitions~\cite{martinez2019inferring,skinner2021estimating,vandermeer2022thermodynamic,harunari2022learn}, to infer thermodynamic quantities such as the EP rate. 
However, most of these studies are concerned with steady states, and only a few of them develop thermodynamic inference methods for nonstationary processes~\cite{shiraishi2015fluctuation,Otsubo2022estimating,Lee2023Multidimensional,Kappler2022Measurement}. Moreover, existing methods for nonstationary processes rely only on nonstationary measurements, thus imposing a relatively high demand for nonstationary data. It is natural to ask whether we can reduce the required nonstationary data by using additional data from other types of measurements.

In this paper, we propose a method of thermodynamic inference for relaxation processes that uses measurements in \textit{tilted equilibrium}, i.e., the equilibrium under the application of external fields to the system. Our approach combines the nonstationary measurement of a few observables with the tilted equilibrium measurement of the same set of observables. From these data, our method allows us to compute the exact value of the minimum EP compatible with the nonstationary data, which constitutes a tight lower bound on the true EP over the relaxation from any intermediate distribution to the final equilibrium. Moreover, if the system satisfies a condition called \textit{realizability condition}, which says that the nonstationary distribution is exactly realized as a tilted equilibrium, our method provides us with additional information about the process: the exact value of the true EP, the instantaneous EP rate, the nonstationary thermodynamic forces, and a constraint on relaxation trajectories. 
Our method applies to arbitrary classical stochastic systems relaxing to equilibrium, including overdamped and underdamped systems, that may have continuous or discrete state spaces.

This paper is organized as follows. 
In Sec.~\ref{sec:setup}, we state the setup and define the problem. 
Section~\ref{sec:inference} is the main section of this paper, where we introduce the tilted equilibrium measurements and establish the inference method of the minimum EP\@. 
In Sec.~\ref{sec:example}, we numerically demonstrate our proposed method with a one-dimensional Brownian particle. 
Section~\ref{sec:additional} develops additional inference methods under the realizability condition.  
In Sec.~\ref{sec:derivation}, we sketch the derivation of the results. 
In the derivation, we develop a Legendre transform at the level of probability distributions from a similar Legendre transform in information geometry~\cite{amari2007methods,amari2016information}. 
Section~\ref{sec:conclusion} concludes the paper.

\section{Nonequilibrium thermodynamics}
\label{sec:setup}

\subsection{Setup}

We consider a stochastic system in contact with a single heat bath at a constant temperature $T$. The system stochastically moves around a continuous state space $\mathcal X\subset\mathbb{R}^{d}$ or a discrete state space $\mathcal X = \{1,\dots,N\}$. 
Let $p(x)$ denote a probability density function for a continuous system and a probability mass function for a discrete case, which satisfies $\int_{\mathcal X}dx\,p(x)=1$. Here and hereafter, $\int_{\mathcal X}dx$ should always be replaced by $\sum_{x\in\mathcal X}$ for discrete cases. 
The state $x$ has energy $\epsilon(x)$, which is assumed to be time independent. The state energy determines the equilibrium distribution $p_\eq\equiv\{p_\eq(x)\}_{x\in\mathcal X}$ with $p_\eq(x)\propto\exp[-\epsilon(x)/\kB T]$, where $\kB$ is the Boltzmann constant.

The system exhibits a relaxation process $\pns(t)\equiv\{\pns(x;t)\}_{x\in\mathcal X}$ for $t\geq 0$, where $t$ is the time variable. For notational convenience, we use tilde ($\tilde{\phantom{\,\cdot\,}}$) for quantities associated with the relaxation process of interest. We assume that the process $\pns(t)$ converges to the equilibrium distribution $\pns(t) \to p_\eq$ as $t\to \infty$. We do not assume any specific time evolution law unless otherwise noted.

\begin{figure}
    \centering
    \includegraphics[width=\hsize,clip]{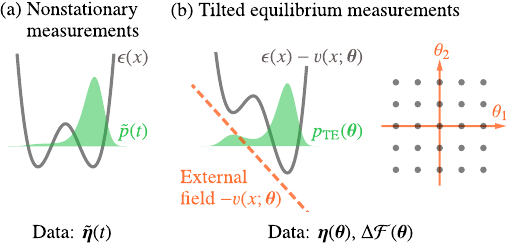}
    \caption{Schematics of the proposed inference method. (a) A system undergoing a relaxation process with a time-independent energy $\epsilon(x)$ and a nonstationary distribution $\pns(t)$. Our method requires only the data of the expectation values of a few observables, $\bmetans(t) = \langle \bm B \rangle_{\pns(t)}$, along the process. 
    (b) To systematically collect data from tilted equilibrium measurements, we apply the external field $v(x;\bm\theta)$ and measure the expectation values of the observables, $\bm\eta(\bm\theta)= \langle \bm B \rangle_{\pTE(\bm\theta)}$, for many $\bm\theta$'s. We also compute the tilted equilibrium free energy $\Delta \F(\bm\theta)$ from the measured values of~$\bm\eta(\bm\theta)$.}
    \label{fig:scheme}
\end{figure}

The fundamental assumption of this paper is that the details of $\pns(x;t)$ are not measurable, but we can measure the expectation values of a few observables $B_{1},\dots,B_{K}$. The number $K\geq 1$ is arbitrary, and it can be a small number such as 2 or 3. An observable $B_{\alpha}\equiv\{B_\alpha(x)\}_{x\in\mathcal X}$ is any real-valued function over $\mathcal X$, and its expectation value over a probability distribution $p$ is $\langle B_{\alpha}\rangle_{p}\coloneqq\int_{\mathcal X}dx\,B_\alpha(x)p(x)$. 
We use the notation $\bm B \equiv (B_1,\dots,B_K)^\trp$ and $\langle \bm B \rangle_p \equiv (\langle B_{1}\rangle_{p},\cdots,\langle B_{K}\rangle_{p})^\trp$.
Without loss of generality, we assume that the $K$ observables are linearly independent of each other. We also assume that the $K$ observables are linearly independent with a constant observable (a trivial observable whose value does not depend on $x$) because such an observable gives no information about the process. We write the set of expectation values measured at time $t$ as 
\begin{equation}
    \bmetans(t)\coloneqq \langle \bm B \rangle _{\pns(t)},
    \label{eq:nonst_data}
\end{equation}
which is the only nonstationary data needed for our proposed method [Fig.~\ref{fig:scheme}(a)].

\subsection{EP and the minimum EP}

The fundamental quantity characterizing the thermodynamic cost of a relaxation process is EP~\cite{seifert2012stochastic}. For systems in contact with a single heat bath, the EP of the relaxation process from $\pns(t)$ to $p_\eq$ is given by $T^{-1}\Delta\H[\pns(t)]$, where
\begin{equation}
    \H[p] \coloneqq\kB T\!\int_{\mathcal X}dx\,p(x)\ln p(x)+\int_{\mathcal X}dx\,\epsilon(x)p(x)
    \label{eq:noneqH}
\end{equation}
and $\Delta\H[p]  \coloneqq \H[p] - \H[p_\eq]$~\cite{seifert2012stochastic,vandenbroeck2015ensemble}. The first term in Eq.~\eqref{eq:noneqH} accounts for the change in the Shannon entropy of the system, while the second term gives the change in the bath entropy due to heat flux. 
Since $T$ is set constant throughout the paper, we abuse terminology and call $\Delta\H[\pns(t)]$ the EP\@.
The EP depends on the details of the probability distribution, and we need $\pns(x;t)$ for all $x$ and all $t$ if we want to compute $\Delta\H[\pns(t)]$ directly from Eq.~\eqref{eq:noneqH}.

Since the measured data $\bmetans(t)$ are not sufficient to determine a single value of $\Delta\H[\pns(t)]$, we follow the general concept in Ref.~\cite{Skinner2021Improved} to focus on the range of EP compatible with the data $\bmetans(t)$. There is no upper limit of the compatible range because hidden (unobserved) degrees of freedom can cause an arbitrarily large EP~\cite{esposito2012stochastic}. On the other hand, there is a lower bound, which we write as 
\begin{equation} 
    \Delta\Hm(\bm\eta)  \coloneqq 
    \min_{\left.q\;\middle|\,\langle\bm B\rangle_{q} =\bm{\eta}\right.} \Delta\H[q],
    \label{eq:minimum_cost}
\end{equation}
for any set of expectation values $\bm\eta$. 
Here, the minimum is taken over all probability distributions $q$ that satisfy $\langle\bm B\rangle_{q} \nobreak=\nobreak\bm{\eta}$, i.e., that are compatible with the set of expectation values $\bm\eta$.
Since $\Delta\H[q]$ is the dissipation of the relaxation process from $q$ to $p_\eq$, any relaxation process that exhibits the set of expectation values $\bm\eta$ at some point in time must dissipate at least $\Delta\Hm(\bm\eta)$ before relaxing to the final equilibrium. For this reason, $\Delta\Hm(\bm\eta)$ is interpreted as the fundamental cost associated with the set of expectation values $\bm\eta$.

\section{Thermodynamic inference}
\label{sec:inference}

\subsection{Tilted equilibrium}
\label{subsec:tilted_eq}

Our main result is a method for calculating the minimum EP $\Delta\Hm(\bm\eta)$ from tilted equilibrium measurements. The method uses a family of external fields parameterized by $\bm\theta=(\theta_1,\dots,\theta_K)^\trp$:
\begin{equation}
    v(x;\bm{\theta})=\sum_{\alpha=1}^K \theta_{\alpha} B_\alpha(x).
    \label{eq:ext_field}
\end{equation}
The external field $v(x;\bm{\theta})$ is the superposition of the fields proportional to the observable $B_{\alpha}$ with an intensity $\theta_{\alpha}$. The external field $v(x;\bm\theta)$ incurs the tilted equilibrium $\pTE(\bm\theta) \equiv \{\pTE(x;\bm\theta)\}_{x\in\mathcal X}$ with 
\begin{equation}
    \pTE(x;\bm{\theta})\propto\exp\qty(-\frac{\epsilon(x)-v(x;\bm{\theta})}{\kB T} )\,.
\end{equation}
Here, we use the convention for the sign of the external field such that it modifies the state energy from $\epsilon(x)$ to $\epsilon(x)-v(x;\bm\theta)$.

We propose the following procedure for collecting tilted equilibrium data to perform the thermodynamic inference systematically [Fig.~\ref{fig:scheme}(b)]:
\begin{enumerate}
    \item Realize the tilted equilibrium distributions for many sets of parameter values $\bm\theta$ and measure the expectation values of the observables $\bm B$,
    \begin{equation}
        \bm\eta(\bm\theta) \coloneqq \langle\bm B \rangle_{\pTE(\bm\theta)},
        \label{eq:def_eta}
    \end{equation}
    for each $\bm\theta$.
    \item Interpolate the measured expectation values to infer the functional dependence $\bm{\eta}(\bm{\theta})$ over a range of $\bm\theta$.
    \item Calculate the tilted equilibrium free energy $\Delta \F(\bm\theta)$ from the data $\bm\eta(\bm\theta)$ as described below.
\end{enumerate}
The tilted equilibrium free energy is defined as 
\begin{equation}
    \F(\bm{\theta}) 
    \coloneqq -\kB T\ln\left[\int_{\mathcal X}dx\exp\left(-\frac{\epsilon(x)-v(x;\bm{\theta})}{\kB T}\right)\right],
    \label{eq:tilted_F}
\end{equation}
and $\Delta\F(\bm\theta) \coloneqq \F(\bm\theta) - \F(\bm 0)$. 
The difference $\Delta\F(\bm\theta)$ is computed from the data $\bm\eta(\bm\theta)$ as (see Sec.~\ref{subsec:derivation_Legendre} for derivation)
\begin{equation}
    \Delta\F(\bm\theta) = - \int_{\bm 0}^{\bm \theta}\bm\eta(\bm\theta')\cdot d\bm\theta',
    \label{eq:line_integral}
\end{equation}
where the line integral is over any curve connecting $\bm\theta'= \bm 0$ and $\bm\theta'=\bm\theta$, and the resulting value does not depend on the intermediate path.

As we show in Sec.~\ref{subsec:derivation_Legendre}, the correspondence from $\bm\theta$ to $\bm\eta$, denoted by $\bm\eta(\bm\theta)$ in Eq.~\eqref{eq:def_eta}, is invertible. We write the inverse function as $\bm\theta(\bm\eta)$, which solves
\begin{equation}
    \langle \bm B \rangle_{\pTE(\bm\theta(\bm\eta))} = \bm\eta 
    \label{eq:def_theta_eta}
\end{equation}
for each $\bm\eta$. In other words, $\bm\theta(\bm\eta)$ is the unique set of parameter values that incurs the set of expectation values $\bm\eta$.
Using the interpolated data of $\bm\eta(\bm\theta)$, one can find $\bm\theta(\bm\eta)$ for a given $\bm\eta$ if the data range covers the given $\bm\eta$. For simplicity, we assume that the data range is large enough to cover any $\bm\eta$ that appears below.

In summary, we have the data of $\bm\eta(\bm\theta)$ and $\Delta\F(\bm\theta)$ from the tilted equilibrium measurements, and when a set of expectation values $\bm\eta$ is given, we can find $\bm\theta(\bm\eta)$ from the tilted equilibrium data.

\subsection{Inference of minimum EP}
\label{subsec:minimal}

In terms of these tilted equilibrium data, the minimum EP in Eq.~\eqref{eq:minimum_cost} is expressed as
\begin{equation}
    \Delta \Hm(\bm\eta)
    = \Delta\F(\bm{\theta}(\bm{\eta}))
    + \sum_{\alpha=1}^K \theta_{\alpha}(\bm{\eta})\,\eta_{\alpha},
    \label{eq:Legendre_main}
\end{equation}
which is the central relation of our inference method.
We can calculate the right-hand side of Eq.~\eqref{eq:Legendre_main} for a given $\bm\eta$ from the tilted equilibrium data. To do so, we first  find $\bm\theta(\bm\eta)$, i.e., the set of parameter values corresponding to the given $\bm\eta$, from the interpolated data $\bm\eta(\bm\theta)$. Then we look up the value of the tilted equilibrium free energy $\Delta\F(\bm{\theta}(\bm\eta))$. Equation~\eqref{eq:Legendre_main} is derived in Sec.~\ref{subsec:derivation_Legendre} and Appendix~\ref{sec:apdx_derivation}.

From Eq.~\eqref{eq:nonst_data} and the definition of $\Delta\Hm(\bm\eta)$ in Eq.~\eqref{eq:minimum_cost}, the true EP of the process for each $t$ is lower bounded as
\begin{subequations}
\label{eq:Legendre_ineq_all}
\begin{align}
    \Delta\H[\pns(t)] 
    &\geq \Delta\Hm(\bmetans(t)) 
    \label{eq:Legendre_ineq_inequality} \\[-0.3\baselineskip]
    &= \Delta \F(\bm{\theta}(\bmetans(t)))  + \sum_{\alpha=1}^K \theta_{\alpha}(\bmetans(t))\,\tilde\eta_{\alpha}(t).
    \label{eq:Legendre_ineq_TE}
\end{align}
\end{subequations}
Equation \eqref{eq:Legendre_ineq_TE} can be calculated by combining the nonstationary data and the tilted equilibrium data. To do so, we first use the tilted equilibrium data to find $\bm\theta(\bmetans(t))$, i.e., the set of parameter values that incurs the same set of expectation values as the nonstationary data $\bmetans(t)$. Then we find the value of the tilted equilibrium free energy $\Delta\F(\bm\theta(\bmetans(t)))$. 

We emphasize that the calculated value of $\Delta \Hm(\bmetans(t))$ is not merely a lower bound on the true EP but a meaningful thermodynamic cost. Indeed, it is the fundamental cost for producing the observed set of expectation values $\bmetans(t)$ in relaxation processes, as discussed below Eq.~\eqref{eq:minimum_cost}.

\begin{figure*}[!bt]
    \centering
    \includegraphics[width=\hsize,clip,page=1]{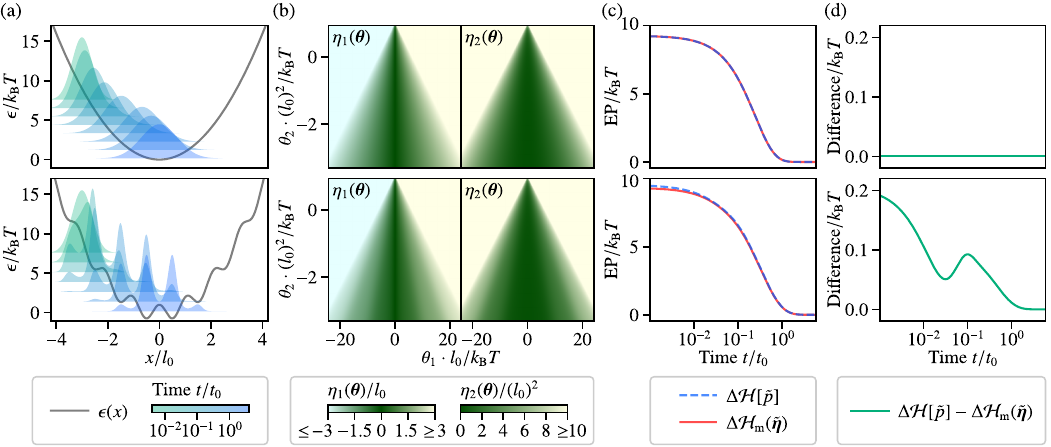}
    \caption{Demonstration of our inference method with a one-dimensional Brownian particle with $\epsilon(x) = ax^2 + b\cos cx$. The upper panels are with $b=0$ (harmonic), and the lower panels are with $b=\kB T$ (ragged). We scale the energy by $\kB T$, the length by $l_0 \coloneqq \sqrt{\kB T/a}$, and the time by $t_0 \coloneqq 1/a\mu$, where $\mu$ is the mobility. See the main text for parameter values. 
    (a) The potential energy $\epsilon(x)$ (gray) and the probability distributions over a relaxation process (blue; vertically shifted). These quantities are assumed to be unmeasurable. (b) The tilted equilibrium measurement measures the expectation values of the observables $(\eta_1(\bm\theta), \eta_2(\bm\theta))$ with varying the external field parameters $(\theta_1,\theta_2)$.
    (c) Inference of the minimum entropy production (EP). The minimum EP $\Delta\Hm(\bmetans(t))$ is obtained from the measurements (red), while the true EP $\Delta\H[\pns(t)]$ is not measurable (blue). 
    (d) The difference between the true EP and the obtained minimum EP, $\Delta\H[\pns(t)]- \Delta\Hm(\bmetans(t))$.}
    \label{fig:example}
\end{figure*}

\subsection{Equality condition and tightness}
\label{subsec:equality}

If the inequality in Eq.~\eqref{eq:Legendre_ineq_inequality} holds with equality, we can use Eq.~\eqref{eq:Legendre_ineq_TE} to calculate the exact value of the true EP from the nonstationary and tilted equilibrium data. As shown in Sec.~\ref{subsec:scaling}, this happens if and only if 
\begin{equation}
    \exists\, \bm\theta,\quad \pns(t) = \pTE (\bm\theta),
    \label{eq:realizability_condition}
\end{equation}
namely, $\pns(t)$ is exactly realizable as a tilted equilibrium. We call the condition in Eq.~\eqref{eq:realizability_condition} the \textit{realizability condition}. We discuss when the realizability condition holds in Sec.~\ref{subsec:sufficient_conditions}.

Apart from the equality condition, we make two remarks about the tightness of the inequality in Eq.~\eqref{eq:Legendre_ineq_inequality}. First, the inequality becomes tighter as we increase $K$ by adding more observables to $\bm B$. This is because increasing $K$ has the effect of narrowing the domain of minimization in Eq.~\eqref{eq:minimum_cost} and raising the minimum EP $\Delta\Hm(\bm\eta)$.

Second, we can sometimes get a tighter bound by considering the whole process at once, assuming that the time evolution is Markovian:
\begin{equation}
    \Delta\H[\pns(t)] 
    =\max_{t'\,|\,t'\geq t}\Delta\H[\pns(t')] 
    \geq\max_{t'\,|\,t'\geq t} \Delta\Hm(\bmetans(t')).
    \label{eq:monotone_correction}
\end{equation}
The first equality is the second law of thermodynamics for the true EP, $-d_{t}\Delta\H[\pns(t)] \geq 0$, where $d_t \equiv d/dt$ is the time derivative, which holds under the Markovian assumption~\cite{seifert2012stochastic,Strasberg2019NonMarkovianity}. The next inequality follows from the inequality in Eq.~\eqref{eq:Legendre_ineq_inequality}. The last side of Eq.~\eqref{eq:monotone_correction} can be computed from the considered data using Eq.~\eqref{eq:Legendre_ineq_TE}, and it gives a tighter bound than Eq.~\eqref{eq:Legendre_ineq_inequality} if $\Delta\Hm(\bmetans(t))$ is not monotonically decreasing in time.

\section{Example: Trapped Brownian particle}
\label{sec:example}

\subsection{Model and required measurements}
\label{subsec:example_setup}

We demonstrate our results with a one-dimensional Brownian particle in a harmonic trapping potential plus a ragged potential,
\begin{equation}
    \epsilon(x)=ax^2+\epsilon_{r}(x),
\end{equation}
where $a>0$ determines the width of the harmonic potential, and $\epsilon_{r}(x)$ is an arbitrary ragged potential. It is a model of an optically trapped particle linked to a biomolecule~\cite{mehta1999single,seifert2012stochastic}, for which $\epsilon_r(x)$ denotes the energy of the biomolecule pulled to a displacement $x$. The time evolution is governed by the Fokker--Planck equation with a uniform mobility $\mu$, $\partial_t  \pns (x;t)=\mu \partial_x [\pns (x;t)\partial_x  \epsilon(x)+\kB T\partial_x \pns(x;t)]$, where $\partial_t \equiv \partial /\partial t$ and $\partial_x \equiv \partial/\partial x $ denote the time and spatial derivatives. 

As an example, we consider the case where we can keep track of the mean and variance of the position $x$ of the particle over relaxation processes, but we cannot reliably get the higher moments due to the limited number of relaxation trajectories obtained from experiments. This assumption corresponds to the choice of two observables, $B_{1}(x)=x$ and $B_2(x)=x^2$, in our framework. To perform the tilted equilibrium measurements, we need to modify the energy to
\begin{equation}
    \epsilon(x)-v(x;\bm{\theta})=(a-\theta_{2})x^2-\theta_{1}x  + \epsilon_r(x),
    \label{eq:example_modified_energy}
\end{equation}
which is still a harmonic potential plus the ragged potential. Therefore, we can realize the tilted equilibrium by modulating the center and the width of the harmonic trapping potential. We need to measure the mean and variance of the position $x$ at many tilted equilibrium distributions, interpolate the data to find  $\bm\eta(\bm\theta)$, and compute $\Delta\F(\bm\theta)$ as described in Sec.~\ref{subsec:tilted_eq}.

If $\epsilon_r(x)=0$ and the initial distribution is Gaussian, the realizability condition [Eq.~\eqref{eq:realizability_condition}] holds. To see this, we first note that $\pns(t)$ is Gaussian for all $t\geq 0$ for a Gaussian initial distribution~\cite{kampen2007stochastic}. Since we can make any harmonic potential by changing $\theta_1$ and $\theta_2$ in Eq.~\eqref{eq:example_modified_energy} for $\epsilon_r(x)=0$, we can realize any Gaussian distribution as a tilted equilibrium. Therefore, the realizability condition is satisfied, and the equality holds in Eq.~\eqref{eq:Legendre_ineq_inequality}, allowing us to extract the exact value of $\Delta\H[\pns(t)]$ from the tilted equilibrium data. In the case of a nonzero ragged potential, the realizability condition no longer holds. Nevertheless, if $\epsilon_r(x)$ is small enough, we expect that $\pns(t)$ is approximately realizable as a tilted equilibrium distribution, and therefore, Eq.~\eqref{eq:Legendre_ineq_TE} gives a fairly accurate estimate of $\Delta\H[\pns(t)]$.

\subsection{Numerical results}
\label{subsec:example_result}

We numerically demonstrate our inference scheme for $\epsilon_r (x) = b \cos cx$. In all calculations, we scale the energy by $\kB T$, the length by $l_0 \coloneqq \sqrt{\kB T/a}$, and the time by $t_0 \coloneqq 1/a\mu$. We set  $b = 0$ (harmonic; the upper panels of Fig.~\ref{fig:example}) or $b=1 \kB T$ (ragged; the lower panels of Fig.~\ref{fig:example}) and $c = 6 (l_0)^{-1}$. The initial distribution is the Gaussian distribution with mean $- 3 l_0$ and variance $0.15 (l_0)^2$ for all calculations. No other parameters are free to choose after scaling the quantities. See also Appendix~\ref{sec:apdx_example} for more details on the numerics.

We plot the potential $\epsilon(x)$ and the time evolution $\pns(t)$, both assumed to be unmeasurable, in Fig.~\ref{fig:example}(a). The ragged system has potential barriers that separate the state space into metastable wells. Figure~\ref{fig:example}(b) shows the tilted equilibrium data $\bm\eta(\bm\theta)$. 
In the ragged system, $\bm\eta(\bm\theta)$ has a steplike feature, reflecting the multiwell structure of the potential.
We sampled $\bm\eta(\bm\theta)$ from sufficiently dense data points over the $\bm\theta$ space, leaving the consideration of sparse data points to future work.

Figure~\ref{fig:example}(c) shows the minimum EP $\Delta\Hm(\bmetans(t))$ compatible with the observed mean and variance, which is calculated using the tilted equilibrium data via Eq.~\eqref{eq:Legendre_ineq_TE}. The true EP $\Delta\H[\pns(t)]$ is also plotted, which is not measurable. The figure shows that the calculated value $\Delta\Hm(\bmetans(t))$ is equal to or less than the true EP $\Delta\H[\pns(t)]$, thus confirming the inequality in Eq.~\eqref{eq:Legendre_ineq_inequality}. Moreover, these two quantities agree exactly for $b=0$, and they are in fairly good agreement for $b = 1 \kB T$ [Fig.~\ref{fig:example}(d)]. The former is expected from the realizability condition, as discussed in Sec.~\ref{subsec:example_setup}\@. The latter, in contrast, is rather surprising because the height of the ragged potential is $2b = 2\kB T$, which is not very small compared with $\kB T$. This example shows that, even if the realizability condition is violated, the minimum EP can be a good approximation of the true EP\@.

\begin{figure}[bt]
    \centering
    \includegraphics{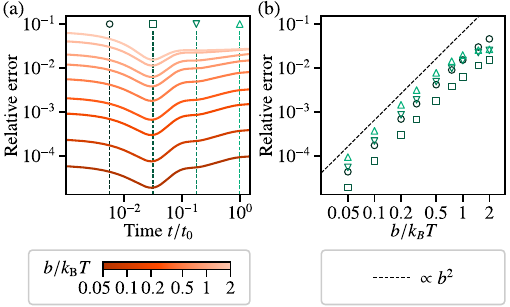}
    \caption{Relative error $\{\Delta\H[\pns(t)]-\Delta\Hm (\bmetans(t)) \} / \Delta\H[\pns(t)] $ between the true entropy production (EP) and the calculated minimum EP\@. (a) Relative error over time, plotted for several values of the height $b$ of the ragged potential. (b) Relative error over $b$, plotted for four time points. Each symbol represents a time point indicated in (a). The relative error grows quadratically in $b$ (dashed line).}
    \label{fig:example_summarize}
\end{figure}

We also present an extensive result over a wide range of $b$. Figure~\ref{fig:example_summarize}(a) shows the relative error, $\bigl\{\Delta\H[\pns(t)] - \Delta\Hm(\bmetans(t)) \bigr\} / \Delta\H[\pns(t)]$, for various values of $b$. The relative error is zero for $b=0$ due to the realizability condition, and it increases with $b$. It does not significantly depend on $t$. As shown in Fig.~\ref{fig:example_summarize}(b), the relative error for a fixed $t$ scales quadratically in $b$. Therefore, the calculated value $\Delta\Hm(\bmetans(t))$ equals the true EP $\Delta\H[\pns(t)]$ up to the first order in $b$. In fact, this quadratic dependence is a generic property, as discussed in Sec.~\ref{subsec:scaling}\@.

\section{Realizability condition and additional inference methods}
\label{sec:additional}

As discussed in Sec.~\ref{subsec:equality}, the realizability condition [Eq.~\eqref{eq:realizability_condition}] ensures that the inequality in Eq.~\eqref{eq:Legendre_ineq_inequality} holds with equality, allowing the inference of the exact value of EP\@. We discuss three situations where the realizability condition is satisfied in Sec.~\ref{subsec:sufficient_conditions}\@. Moreover, assuming the realizability condition, we can extract additional information about the relaxation process from the tilted equilibrium measurements, including the EP rate, the thermodynamic force, and a constraint on relaxation paths. We present these additional inference methods in Secs.~\ref{subsec:additional_EPR}--\ref{subsec:additional_L}\@.

\subsection{Sufficient conditions for the realizability condition}
\label{subsec:sufficient_conditions}

There are several physically natural situations in which we can ensure the realizability condition (see Appendix~\ref{sec:apdx_realizability condition} for details). 
The first situation is the existence of a time-scale separation: We can ensure the realizability condition except for the initial fast relaxation if the states are lumped into groups, the Markovian transition rates within a group are sufficiently larger than the rates between different groups, and we can track the total probability of each group as the observables.
The second situation is when the system admits a symmetry: The realizability condition holds if the system energy and the time evolution law are symmetric under some permutations of the discrete states, the initial condition has the same symmetry, and we can track all observables obeying the symmetry. 
Another situation is when the process is Gaussian: We can ensure realizability condition if the state energy is a (possibly multivariate) harmonic potential, the time evolution is governed by the Fokker--Planck equation with a uniform mobility, the initial distribution is Gaussian, and we can track the mean and covariance matrix as the observables. This third situation has been demonstrated in the example (Sec.~\ref{sec:example})\@. 
Note that each of these conditions jointly concerns the system energy, the time evolution law, the initial condition, and the choice of observables.

If we have enough information about the system to ensure that the system satisfies one of these sufficient conditions for the realizability condition, we can obtain the exact value of the true EP from Eq.~\eqref{eq:Legendre_ineq_TE}, and we can obtain additional information by the methods described below in Secs.~\ref{subsec:additional_EPR}--\ref{subsec:additional_L}\@. 
Alternatively, if we can expect the system to approximately satisfy one of these sufficient conditions, we expect $\pns(t)$ to be approximately realized as a tilted equilibrium distribution. In this case, $\Delta\Hm(\bmetans(t))$ will be a good approximation of $\Delta\H[\pns(t)]$, and the additional inference methods below will provide reasonable estimates of the additional quantities.

\subsection{EP rate}
\label{subsec:additional_EPR}

The first quantity obtained under the realizability condition is the EP rate, $\epr(t) \coloneqq -T^{-1}d_t  \Delta\H[\pns(t)] $. Under the realizability condition, we have the equality $\Delta \H[\pns (t)] = \Delta \Hm(\bmetans(t))$, and thus, we can obtain the EP rate by simply differentiating $ \Delta \Hm(\bmetans(t))$, which is calculated from Eq.~\eqref{eq:Legendre_ineq_TE}. Alternatively, we have the following equality (see Sec.~\ref{subsec:derivation_Legendre} for derivation):
\begin{equation}
    \epr(t)
    = -\frac 1T d_t \Delta \Hm(\bmetans(t)) =  - \frac 1T \sum_{\alpha=1}^K
    \theta_{\alpha}(\bmetans(t)) \, d_t  \tilde\eta_{\alpha}(t).
    \label{eq:epr_decomp}
\end{equation}
The right-hand side of this equation is calculated by differentiating the nonstationary data $\bmetans(t)$ and finding $\bm\theta(\bmetans(t))$ from the tilted equilibrium data $\bm\eta(\bm\theta)$. 

Equation~\eqref{eq:epr_decomp} also provides a decomposition of the EP rate into the dissipation due to the change in the expectation value of each observable. From this equation, we can regard $-T^{-1}\theta_{\alpha}(\bmetans(t))$ as the thermodynamic force conjugate to the probability flux incurring the change in the expectation values $d_t \tilde\eta_{\alpha}(t)$. 

\begin{figure}
    \centering
    \includegraphics[width=\hsize,clip]{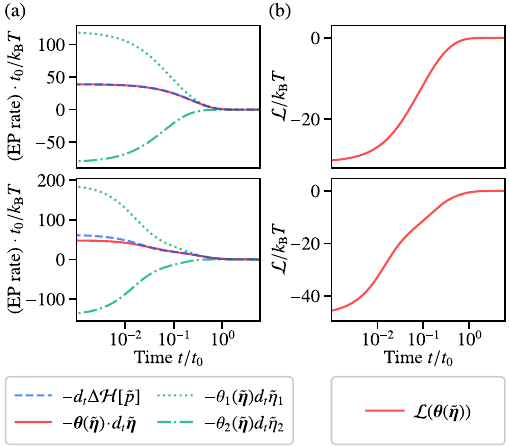}
    \caption{Additional inference under the realizability condition demonstrated with the same systems as in Fig.~\ref{fig:example}. The harmonic system (upper panels) satisfies the realizability condition, while the ragged system (lower panels) satisfies it only approximately. 
    (a) Inference of the entropy production (EP) rate. The true EP rate $T\epr(t) = -d_t \Delta\H[\pns(t)]$ (blue) and the obtained estimate $-\bm\theta(\bmetans(t)) \cdot d_t \bmetans(t)$ (red) are in exact agreement under the realizability condition. We also plot the decomposed values $-\theta_\alpha(\bmetans(t)) d_t \tilde\eta_\alpha(t)$ (green).
    (b) Constraint on Markovian time evolution. The function $\mathcal L(\bm\theta(\bmetans(t)))$ increases monotonically with time under the realizability condition.}
    \label{fig:example_additional}
\end{figure}

In Fig.~\ref{fig:example_additional}(a), we demonstrate the inference of the EP rate from Eq.~\eqref{eq:epr_decomp} for the same example systems as in Sec.~\ref{sec:example}\@. Without a ragged potential (the upper panel), the realizability condition is exactly satisfied, and therefore the right-hand side of Eq.~\eqref{eq:epr_decomp} gives the exact value of $\epr(t)$. With a nonzero ragged potential (the lower panel), the realizability condition is approximately satisfied, and indeed the right-hand side of Eq.~\eqref{eq:epr_decomp} gives a good estimate of the EP rate.

\subsection{Thermodynamic force}
\label{subsec:additional_force}

The second quantity obtained under the realizability condition is the thermodynamic force (affinity) over the state space. 
For discrete-state systems, the thermodynamic force from state $y$ to $x$ at time $t$ is given by~\cite{vandenbroeck2015ensemble}
\begin{equation}
    \tilde{f}(x,y;t) \coloneqq - \frac{\epsilon(x)-\epsilon(y)}{T} - \kB \ln \frac{\pns(x;t) }{ \pns(y;t)}.
    \label{eq:thermo_force_def}
\end{equation} 
The thermodynamic force is related to the EP rate by $\epr (t) = \frac{1}{2} \sum_{x,y} \tilde{\jmath}(x,y;t)\, \tilde f(x,y;t)$, where $\tilde{\jmath}(x,y;t)$ is the net probability flux from state $y$ to $x$ at time $t$~\cite{vandenbroeck2015ensemble}. Thus, the thermodynamic force quantifies the EP rate due to the probability flux from $y$ to $x$. Under the realizability condition, we can calculate the thermodynamic force $\tilde{f}(x,y;t)$ from the considered data as (see Appendix \ref{subsec:apdx_thermo_force} for derivation)
\begin{equation}
    \tilde{f}(x,y;t)= - \frac{1}{T} \sum_{\alpha=1}^K \theta_{\alpha}(\bmetans(t))\, [B_{\alpha}(x)-B_\alpha(y)].
    \label{eq:thermo_force}
\end{equation}
The right-hand side can be calculated from the nonstationary data $\bmetans(t)$ and the tilted equilibrium data $\bm\eta(\bm\theta)$, assuming that we know the values of the observables $\bm B(x)$ in each state.

For continuous-state systems, the thermodynamic force at $x\in \mathbb R^d$ is defined as a continuous version of Eq.~\eqref{eq:thermo_force_def}~\cite{vandenbroeck2010three}, $\tilde{f}(x;t) \coloneqq - T^{-1} \nabla \epsilon(x) - \kB \nabla \ln \pns(x;t)$, which is a $d$-di\-men\-sion\-al vector. It is related to the EP rate by $\epr (t) = \int_{\mathcal X} dx \, \tilde{\jmath}(x;t) \cdot \tilde{f}(x;t)$, where $\tilde{\jmath}(x;t)$ is the $d$-dimensional probability current at time $t$.  We can similarly calculate the thermodynamic force from the data as $\tilde{f}(x;t) = - T^{-1} \sum_{\alpha=1}^K \theta_\alpha(\bmetans(t)) \nabla B_\alpha (x)$ under the realizability condition.

\subsection{Constraint on the time evolution}
\label{subsec:additional_L}

The final piece of information obtained from tilted equilibrium measurements is a constraint on the time evolution.
Let us introduce a function of the external field parameters~$\bm\theta$:
\begin{equation}
    \mathcal L(\bm\theta) \coloneqq \Delta \F(\bm{\theta}) + \sum_{\alpha=1}^K \theta_\alpha \langle 
    B_\alpha  \rangle_{p_\eq} .
    \label{eq:monotonicity_function}
\end{equation}
Then the time evolution of the observables $\bmetans(t)$ must satisfy
\begin{equation}
    d_t \mathcal L(\bm\theta(\bmetans(t))) \geq 0,
    \label{eq:monotonicity}
\end{equation}
assuming a Markovian time evolution in addition to the realizability condition (see Appendix \ref{subsec:apdx_monotonicity} for derivation).
This restriction on the possible time trajectories is different from and complementary to the second law of thermodynamics, $-d_t \Delta\H[\pns(t)] \geq 0$.
Note that $\langle B_\alpha \rangle _{p_\eq}$ is independent of $\bm\theta$, and the second term of $\mathcal L(\bm\theta)$ is linear in $\bm\theta$. If we shift the definition of each observable by a constant so that $\langle B_\alpha \rangle_{p_\eq}=0$, $\mathcal L(\bm\theta)$ coincides with $\Delta\F(\bm\theta)$.

In Fig.~\ref{fig:example_additional}(b), we demonstrate the monotonicity in Eq.~\eqref{eq:monotonicity} with the same example system as above. For the harmonic system (the upper panel), the realizability condition is satisfied, and the function $\mathcal L(\bm\theta)$ is indeed monotonically increasing. For the ragged  system (the lower panel), the realizability condition holds only approximately, but $\mathcal L(\bm\theta)$ is still monotonically increasing.

\section{Derivation}
\label{sec:derivation}

\subsection{Derivation of the central relation \texorpdfstring{[Eq.~\eqref{eq:Legendre_main}]}{}}
\label{subsec:derivation_Legendre}

The derivation of Eq.~\eqref{eq:Legendre_main} involves two nontrivial tasks: expressing the minimum EP in Eq.~\eqref{eq:minimum_cost} in a closed form and relating the minimum EP to the tilted equilibrium quantities. We sketch how these two tasks are accomplished, leaving the detailed calculation to Appendix \ref{sec:apdx_derivation}\@.

To express the minimum EP in a closed form, we first find the following relation for any set of expectation values $\bm\eta$ and any distribution $q$ that satisfies $\langle \bm B \rangle_q = \bm\eta$:
\begin{equation}
    \Delta \H[q] = \kB T \DKL [ q \| \pTE (\bm\theta(\bm\eta)) ] + \Delta \H [\pTE (\bm\theta(\bm\eta))] ,
    \label{eq:pythagoras} 
\end{equation}
where $\DKL[p_1\| p_2] = \int_{\mathcal X} dx \,p_1(x) \ln [p_1(x)/p_2(x)]$ is the Kullback--Leibler (KL) divergence~\cite{kullback1951information}. 
Taking the minimum of both sides of Eq.~\eqref{eq:pythagoras} with respect to $q$ that satisfies $\langle \bm B \rangle_q = \bm\eta$ for a fixed $\bm\eta$, the left-hand side reduces to the definition of $\Delta\Hm(\bm\eta)$ in Eq.~\eqref{eq:minimum_cost}. The minimum of the right-hand side is achieved at $q=\pTE(\bm\theta(\bm\eta))$ since the KL divergence $D[p_1\| p_2]$ is nonnegative, and it is zero if and only if $p_1=p_2$. Thus we have
\begin{equation}
    \Delta\Hm(\bm\eta) = \Delta\H[\pTE (\bm\theta(\bm\eta))].
    \label{eq:minimum_cost_closed}
\end{equation}
This successfully expresses the minimum EP in a concrete form.

The other element of the proof is a Legendre duality over probability distributions. Using the expression of $\Delta\Hm(\bm\eta)$ in Eq.~\eqref{eq:minimum_cost_closed}, we can prove that the two functions $-\Delta\F(\bm{\theta})$ and $\Delta\Hm(\bm{\eta})$ are strictly convex and connected by a Legendre transform: 
\begin{subequations}
\label{eq:Legendre_math}
\begin{gather}
    \eta_{\alpha}(\bm{\theta})=-\frac{\partial\Delta\F(\bm\theta)}{\partial\theta_{\alpha}}, \quad
    \theta_{\alpha}(\bm{\eta})=\frac{\partial\Delta\Hm(\bm\eta)}{\partial\eta_{\alpha}},
    \label{eq:Legendre_math_a} \\
    \Delta\Hm(\bm{\eta})=\Delta\F(\bm{\theta}(\bm{\eta})) + \bm{\theta}(\bm{\eta})\cdot\bm{\eta},\hspace{1.8em}
    \label{eq:Legendre_math_b}
\end{gather}
\end{subequations}
where the correspondences $\bm{\theta}(\bm{\eta})$ and $\bm{\eta}(\bm{\theta})$ are the same as those already defined in Sec.~\ref{subsec:tilted_eq}\@. Equation~\eqref{eq:Legendre_math_b} is identical to our central relation in Eq.~\eqref{eq:Legendre_main}. Equation~\eqref{eq:Legendre_math_a} ensures that the correspondence between $\bm\theta$ and $\bm\eta$ is one-to-one, and thus the solution to Eq.~\eqref{eq:def_theta_eta} is unique. Equation~\eqref{eq:Legendre_math_a} also proves the expression of $\Delta\F(\bm\theta)$ as a line integral in Eq.~\eqref{eq:line_integral}, as well as the expression of the EP rate in Eq.~\eqref{eq:epr_decomp}.

\subsection{Equality condition and scaling of the error}
\label{subsec:scaling}

We prove that the realizability condition [Eq. \eqref{eq:realizability_condition}] is the equality condition of the inequality in Eq.~\eqref{eq:Legendre_ineq_inequality}. Inserting $q = \pns(t)$ and $\bm\eta = \bmetans(t)$ into Eqs.~\eqref{eq:pythagoras} and~\eqref{eq:minimum_cost_closed}, we find that the difference between the two sides of the inequality in Eq.~\eqref{eq:Legendre_ineq_inequality} is given by
\begin{equation}
    \Delta \H[\pns(t)] - \Delta \Hm(\bmetans(t)) = \kB T \DKL[\pns(t) \| \pTE(\bm\theta(\bmetans(t)))].
    \label{eq:error_explicit}
\end{equation}
Since the KL divergence is zero if and only if the two arguments are equal, this difference vanishes if and only if $\pns(t) = \pTE(\bm\theta(\bmetans(t)))$. If this happens, then obviously the realizability condition is satisfied. Conversely, if the realizability condition holds, then the set of parameter values $\bm\theta$ in Eq.~\eqref{eq:realizability_condition} must satisfy $\langle \bm B \rangle_{\pTE(\bm\theta)} = \langle \bm B \rangle_{\pns(t)} = \bmetans(t) $, and therefore, it is given by $\bm\theta = \bm\theta(\bmetans(t))$. Thus, we have $\pns(t) = \pTE(\bm\theta(\bmetans(t)))$, and  Eq.~\eqref{eq:error_explicit} vanishes.

We can also show that the difference in Eq.~\eqref{eq:error_explicit} scales quadratically with the magnitude of the violation of the realizability condition, which explains the observed behavior in Fig.~\ref{fig:example_summarize}(b). Consider a reference system that satisfies the realizability condition and another perturbed system with a slightly different energy, time evolution equation, initial condition, or set of observables. We use $\lambda$ for the magnitude of any of these perturbations. 
In Appendix~\ref{sec:apdx_perturbation}, we show that the perturbed system generically obeys the scaling $\pns(t) - \pTE(\bm\theta(\bmetans(t))) = O(\lambda)$, assuming a Markovian time evolution and some mild conditions. Combined with the expansion of the KL divergence between two close distributions, $\DKL[p_1\|p_2] \simeq \frac 12 \int_{\mathcal X}dx \,[p_1(x) - p_2(x)]^2/p_1(x) $ in the leading order of $p_1 - p_2$~\cite{amari2016information}, we can see that the error term scales as $\DKL[ \tilde p(t) \| \pTE(\bm\theta(\bmetans(t)))] = O(\lambda^2)$. Combined with Eq.~\eqref{eq:error_explicit}, we conclude 
\begin{equation}
    \Delta\H[\tilde p(t)] = \Delta\Hm(\bmetans(t)) + O(\lambda^2).
\end{equation}
Therefore, our method gives the true EP up to the first order in the magnitude of the violation of the realizability condition.

\section{Discussion}
\label{sec:conclusion}

In this paper, we have developed a method of thermodynamic inference that uses tilted equilibrium measurements. The method enables us to obtain the exact value of the minimum EP $\Delta\Hm(\bm\eta)$ compatible with the observed set of expectation values $\bm\eta$. This method applies to any classical stochastic system that relaxes to equilibrium with any choice of observables. Furthermore, if we have enough information about the system to ensure that the realizability condition holds, or at least that the realizability condition is approximately satisfied, we can extract the true EP, the EP rate with its decomposition, the thermodynamic force, and a constraint on relaxation paths.

Compared with existing methods of thermodynamic inference for nonstationary processes~\cite{shiraishi2015fluctuation,Otsubo2022estimating,Lee2023Multidimensional,Kappler2022Measurement}, our approach significantly reduces the demand for nonstationary measurements. Our method requires only the expectation values of a few arbitrary observables, which is insufficient for any of these existing methods. This reduced demand is achieved at the expense of tilted equilibrium measurements.
Therefore, our method will be useful when one cannot practically collect sufficiently many trajectories of a relaxation process to infer EP only from nonstationary data using previously proposed methods, but one can freely apply a few types of external fields to the system. This would include both experiments and numerical simulations.

Our method generally provides only the lower bound of EP, but it gives the optimal lower bound in the sense that there exists a distribution $\pns(t)$ that saturates the inequality in Eq.~\eqref{eq:Legendre_ineq_inequality} for any $\bmetans(t)$ since $\Delta\Hm(\bm\eta)$ is defined by a minimization in Eq.~\eqref{eq:minimum_cost}. In other words, given only $\bmetans(t)$ for each $t$ separately as nonstationary data, our lower bound is the best possible. Moreover, our lower bound $\Delta\Hm(\bm\eta)$ is a meaningful thermodynamic cost since it is interpreted as the minimum cost required to realize the set of expectation values $\bm\eta$, as discussed below Eq.~\eqref{eq:minimum_cost}. This is in contrast with other lower bounds of EP that involve only a few observable values, such as from thermodynamic uncertainty relations for nonstationary processes~\cite{Dechant2018CurrentFluctuations,Liu2020ThermodynamicUncertainty,Koyuk2020Thermodynamic}. These lower bounds are meaningful as statistical quantities, such as the precision of a current observable, but they do not admit interpretations as thermodynamic costs in general.

Our results open an avenue of thermodynamic inference: inference for nonstationary processes based on static measurements. We leave several directions open for future work.
First, our method assumes that the set of measurable observables and the set of available external fields are both given by $\bm B$. However, these two sets are often different in natural situations. Considering this difference is important to make our method more useful. 
Second, our method is exact in the sense that it provides the exact value of the minimum EP $\Delta \Hm(\bm\eta)$. Approximating the minimum EP with a smaller amount of tilted equilibrium data is an interesting direction.
Finally, it is essential to extend the applicability of our approach. As discussed in Appendix \ref{subsec:apdx_generalized_setup_chempot}, our results can be easily extended to systems with internal entropy of states~\cite{esposito2012stochastic} and exchange of particles with a single particle reservoir. On the other hand, extending our results to systems with multiple baths is nontrivial. This extension is possible on a formal (mathematical) level by considering tilted nonequilibrium stationary distributions (see Appendix \ref{subsec:apdx_generalized_setup_NESS}), but making it practical is left to future work. Extending our approach to driven systems is also a nontrivial and important direction.

From a mathematical point of view, the Legendre transform we have introduced in Eq.~\eqref{eq:Legendre_math} is at the level of probability distributions, and it does not rely on asymptotics. Such a microscopic Legendre transform has previously appeared in other fields such as information geometry~\cite{amari2007methods,amari2016information} and the foundations of statistical mechanics~\cite{commons2021duality}, and we have formulated the Legendre transform in Eq.~\eqref{eq:Legendre_math} by borrowing ideas from information geometry and replacing information theoretic quantities with thermodynamic ones, such as the Shannon entropy with the thermodynamic EP\@. This formulation extends the existing connections between information geometry and thermodynamics~\cite{weinhold1975metric,ruppeiner1979thermodynamics,crooks2007measuring,ito2018prl}. Detailed explorations of this geometric picture of stochastic thermodynamics is left for future work.

\acknowledgements

We thank Kohei Yoshimura, Shin-ichi Sasa, Zhiyue Lu, Koki Shiraishi, Artemy Kolchinsky, Ken Hiura, and Ryuna Nagayama for fruitful discussions about the current and/or previous versions of this paper.
N.~O.~is supported by JSPS KAKENHI Grant No.\ 23KJ0732. 
S.~I.~is supported by JSPS KAKENHI Grants No.\ 19H05796, No.\ 21H01560, No.\ 22H01141, and No.\ 23H00467, JST Presto Grant No.\ JPMJPR18M2, JST ERATO Grant No.\ JPMJER2302, and UTEC-UTokyo FSI Research Grant Program.


\appendix

\titleformat{\section}[block]{\centering\fontsize{10.5pt}{\baselineskip}\selectfont\bfseries}{Appendix \Alph{section}:}{10.5pt}{}

\section{Sufficient conditions for the realizability condition}
\label{sec:apdx_realizability condition}

We discuss three situations in which we can ensure the realizability condition in Eq.~\eqref{eq:realizability_condition}. A similar discussion based on information geometry that relates time evolutions of stochastic systems to a constrained set of probability distributions is found in Ref.~\cite{kolchinsky2021prx}.

\subsection{Time-scale separation}
The first situation is the existence of a time-scale separation. Consider that the states are grouped (coarse-grained) into $K+1$ disjoint sets of states $G_0, G_1,\dots,G_K$. We assume the time-scale separation, i.e., that the Markovian transitions between two states in one group are much faster than the transitions between two states in different groups~\cite{rahav2007fluctuation,esposito2012stochastic,Bo2014EntropyProduction}. We further assume that we can keep track of the expectation values of the observables $\chi_0,\chi_1,\dots,\chi_K$, where $\chi_\alpha(x)$ is defined as $\chi_\alpha(x) = 1 $ for $x\in G_\alpha$ and $\chi_\alpha(x) = 0$ for $x \notin G_\alpha$. 
The expectation value $\langle \chi_\alpha \rangle_{\pns(t)}$ gives the total probability of the $\alpha$th group at time $t$. 
The observables satisfy $\sum_{\alpha=0}^K \chi_\alpha(x) = 1$ for all $x$. 
To connect this setup to our formulation, we choose $B_\alpha = \chi_\alpha$ for $\alpha=1,\dots,K$ as the observables. Here, we exclude $\alpha=0$ so that the set of observables $\bm B$ is linearly independent with the constant observable.

Under these assumptions, we prove the realizability condition except for the initial fast relaxation. 
For this purpose, we introduce the coarse-grained probability $\tilde Q(\alpha; t) \coloneqq \langle \chi_\alpha \rangle_{\pns(t)}$ and its equilibrium value $Q_\eq(\alpha) \coloneqq \langle \chi_\alpha \rangle_{p_\eq}$. The time-scale separation implies that the conditional probability within each group, $\pns(x;t)/\tilde Q(\alpha;t)$ for $x \in G_\alpha$, rapidly relaxes to the conditional canonical distribution $p_\eq(x)/Q_\eq(\alpha)$~\cite{rahav2007fluctuation,esposito2012stochastic,Bo2014EntropyProduction}. Thus, the nonstationary probability distribution has the form of
\begin{equation}
    \pns(x;t) = \frac{p_\eq(x)}{Q_\eq(\alpha)} \tilde Q(\alpha; t)\quad (x\in G_\alpha),
    \label{cn_separation}
\end{equation}
except for the initial fast relaxation. Using Eq.~\eqref{cn_separation} and $p_\eq(x) = \exp\{ -[\epsilon(x) - F_\eq ]/\kB T\}$, where $F_\eq$ is the equilibrium free energy, we have
\begin{equation}
    \kB T \ln \pns(x;t) = -\epsilon(x) + F_\eq +  \kB T \sum_{\alpha=0}^K \chi_\alpha(x) \ln \frac{\tilde Q(\alpha; t)}{Q_\eq(\alpha)}.
    \label{cn_separation2}
\end{equation}
By inserting $\chi_\alpha(x) = B_\alpha(x)$ for $\alpha\geq 1$ and $\chi_0(x) = 1- \sum_{\alpha=1}^K B_\alpha(x)$ into Eq.~\eqref{cn_separation2}, we can rearrange Eq.~\eqref{cn_separation2} as
\begin{equation}
    \kB T \ln \pns(x;t) =  - \epsilon(x) +  \sum_{\alpha=1}^K \theta_\alpha  B_\alpha(x) + \mathrm{const.},
\end{equation}
with a set of real numbers $(\theta_\alpha)_{\alpha=1}^K$, where the constant term is independent of $x$. Rearranging gives
\begin{equation}
     \pns(x;t) \propto \exp\qty(-\frac{\epsilon(x) - \sum_{\alpha=1}^K \theta_\alpha  B_\alpha(x)}{\kB T}) .
\end{equation}
Thus, $\pns(x;t)$ is realized as a tilted equilibrium with an external field of the form of Eq.~\eqref{eq:ext_field}, which proves the realizability condition.

Similarly, we can expect the realizability condition to hold approximately if the transitions within a group are faster than the transitions between different groups, but their time scales are not sufficiently separated.

\subsection{Symmetry}

Another situation is when the system and the initial distribution obey a symmetry~\cite{kolchinsky2021prx}. Focusing on a discrete space $\mathcal X=\{1,\dots,N\}$, we consider a symmetry expressed by a permutation group $\mathcal G$ on $\mathcal X$, whose element $g \in \mathcal G$ is a bijection from $\mathcal X$ to itself. We assume that the state energy is invariant under the permutations, $\epsilon(x) = \epsilon(g(x))$ for all $g\in\mathcal G$ and all $x\in\mathcal X$. 
We also assume that the time evolution law obeys the symmetry. For example, if the time evolution is Markovian and given by $\partial_t \pns(x;t) = \sum_y [W(x,y)\pns(y;t) - W(y,x)\pns(x;t)]$, where $W(x,y)$ is the transition rate from $y$ to $x$, then we assume $W(x,y)=W(g(x),g(y))$ for all $g\in \mathcal G$ and all $x,y\in\mathcal X$.
We also impose the symmetry on the initial distribution $\pns(x;0) = \pns(g(x);0)$. We assume that we can keep track of the expectation values of all observables that are invariant under the permutations. In other words, the observables $B_1(x), \dots, B_K(x)$ and the constant observable form a basis of the linear subspace $\{ w \in \mathbb R^N \,\vert\, w(x) = w(g(x)), \forall g \in \mathcal{G}, \allowbreak \forall x \in \mathcal{X} \}$ of $\mathbb R^N$. 

Under these conditions, we prove the realizability condition in Eq.~\eqref{eq:realizability_condition} for all $t\geq 0$. First, by symmetry considerations, the nonstationary distribution obeys the symmetry for all $t>0$:
\begin{equation}
    \pns(x;t) = \pns(g(x);t).
    \label{cn_symmetry}
\end{equation}
Therefore, the vector $\{\kB T\ln \pns(x;t) + \epsilon(x)\}{}_{x\in\mathcal X}$ has the same symmetry, and thus, it can be expanded in terms of $B_1,\dots, B_K$ and the constant observable as 
\begin{equation}
    \kB T \ln \pns(x;t) + \epsilon(x) = \theta_0 + \sum_{\alpha=1}^K \theta_\alpha B_\alpha(x)
    \label{cn_symmetry_2}
\end{equation}
with a set of real numbers $(\theta_\alpha)_{\alpha=0}^K$.
Rearranging Eq.~\eqref{cn_symmetry_2} gives
\begin{equation}
    \pns(x;t) \propto \exp\qty(- \frac{\epsilon(x) - \sum_{\alpha=1}^K \theta_\alpha  B_\alpha(x) }{\kB T} ).
    \label{cn_expsymmetry}
\end{equation}
This is the form of the equilibrium distribution under the external field $\sum_\alpha \theta_\alpha  B_\alpha (x)$. Therefore, $\pns(t)$ is realizable as a tilted equilibrium, and the realizability condition holds.

Similarly, the realizability condition is expected to hold approximately if the symmetry does not hold perfectly but holds approximately.

\subsection{Harmonic potential}

The third situation is when the system has a (possibly multivariate) harmonic potential. More precisely,  focusing on a continuous state space $x \equiv (x_1,\dots,x_d)^\trp \in \mathcal X = \mathbb R^d$, we assume that the potential is harmonic, $\epsilon(x) = x^\trp a x + b^\trp x $, where $a$ is a $d\times d$ positive-definite symmetric matrix, and $b$ is a $d$-dimensional vector. We also assume that the time evolution is given by the Fokker--Planck equation,
\begin{equation}
    \partial_t \pns(x;t) = \nabla\cdot \qty[  \pns(x;t) \mu \nabla \epsilon(x)  + \kB T \mu \nabla \pns(x;t) ],
    \label{cn_harmonic_FP}
\end{equation}
where $\mu$ is a $d\times d$ positive-definite symmetric mobility tensor, with a Gaussian initial distribution. As for the observables, we assume that we can keep track of the mean and the covariance matrix. This amounts to choosing $x_1,\ldots, x_d,\allowbreak (x_1)^2, \ldots (x_d)^2, \allowbreak x_1x_2, x_1x_3,\ldots, x_{d-1}x_d$ as the observables $B_1(x),\dots, B_{2d + d(d-1)/2}(x)$.

Under this setup, we prove the realizability condition in Eq.~\eqref{eq:realizability_condition} for all $t \geq0$. First, the solution of Eq.~\eqref{cn_harmonic_FP} is Gaussian for all $t\geq 0$~\citep[Sec.~VIII 6,][]{kampen2007stochastic}, and therefore, it can be written in a form of
\begin{equation}
    \tilde p(x;t) \propto \exp \qty(  - \frac{x^\trp \tilde a(t) x + \tilde b(t)^\trp x }{\kB T} )
    \label{cn_Gaussian}
\end{equation}
using a $d\times d$ symmetric matrix $\tilde a(t)$ and a $d$-dimensional vector $\tilde b(t)$. We can rewrite Eq.~\eqref{cn_Gaussian} as 
\begin{equation}
    \tilde p(x;t) \propto \exp \qty(  - \frac{\epsilon(x) + x^\trp [\tilde a(t) -a] x + [\tilde b(t) - b] ^\trp x }{\kB T} )\,,
\end{equation}
and further rewrite $x^\trp [\tilde a(t) -a] x + [\tilde b(t) - b] ^\trp x $ as a linear combination of the observables. This shows that $\pns(x;t)$ is realized as a tilted equilibrium distribution with an external field of the form $\sum_\alpha \theta_\alpha B_\alpha(x) $, and thus, the realizability condition is satisfied.

Even if the initial distribution is not Gaussian, the realizability condition holds asymptotically after the initial fast transition. Restricting to a one-dimensional case for simplicity, in which $a$, $b$, and $\mu$ are scalar, we write the solution of the Fokker--Planck equation in terms of the changes in the cumulants. The deviation of the $n$th cumulant ($n = 1,2,\dots$) from its equilibrium value decays exponentially in time as $\exp (-2n \mu a t)$~\cite{kampen2007stochastic}, and the equilibrium values are zero for $n\geq 3$. Thus, only the first two cumulants remain significant after the initial fast transition, which means that the distribution is close to Gaussian, and the realizability condition holds asymptotically. 

We can expect that the realizability condition holds approximately when the potential is close to but not exactly Gaussian. We explore this situation in the example in Sec.~\ref{sec:example}.

\section{Detailed derivation of the results}
\label{sec:apdx_derivation}

In Appendixes~\ref{subsec:apdx_tiltedfreeenergy}--\ref{subsec:apdx_Legendre}, we follow the steps outlined in Sec.~\ref{subsec:derivation_Legendre} to derive the inference of the minimum EP~[Eq.~\eqref{eq:Legendre_main}]. 
In Appendixes~\ref{subsec:apdx_thermo_force} and \ref{subsec:apdx_monotonicity}, we derive the additional inference schemes described in Sec.~\ref{sec:additional}.

\subsection{KL divergence}
\label{subsec:apdx_tiltedfreeenergy}

We start the derivation by relating the thermodynamic quantities to the KL divergence. For any probability distribution $p$, 
\begin{align}
    &\kB T \DKL [p \| p_\eq] 
    \notag \\
    &\quad = \kB T \int_{\mathcal X} dx\, p(x) \ln \frac{p(x)}{p_\eq(x)} 
    \notag \\
    &\quad = \kB T \int_{\mathcal X} dx\, p(x) \ln p(x) + \int_{\mathcal X} dx\, p(x) [\epsilon(x) - F_\eq ]
    \notag \\
    &\quad = \H[p] - F_\eq,
    \label{pr_KL_H_calculation}
\end{align}
where $F_\eq \equiv \mathcal F(\bm 0)$ is the equilibrium free energy at $p_\eq$, and we have used $p_\eq(x) = \exp\{ - [\epsilon(x) - F_\eq] /\kB T\}$. For $p= p_\eq$, Eq.~\eqref{pr_KL_H_calculation} gives $\H[p_\eq] = F_\eq$. Thus, we obtain
\begin{equation}
    \Delta\H[p] =  \kB T \DKL [p \| p_\eq] ,
    \label{pr_KL_H} 
\end{equation}
which is a well-known expression of the EP~\cite{Vaikuntanathan2009DissipationLag,takara2010generalization,esposito2011EPL}.

Next, we rewrite the tilted equilibrium free energy in terms of the KL divergence. For two arbitrary sets of external field parameters, $\bm\theta$ and $\bm\theta'$, we obtain
\begin{align}
    &\kB T \DKL [\pTE(\bm\theta) \| \pTE(\bm\theta')] 
    \notag \\
    & \quad = \kB T \int_{\mathcal X} dx\, \pTE(x;\bm\theta) \ln\frac{\pTE(x;\bm\theta)}{\pTE(x;\bm\theta')} 
    \notag \\
    & \quad = \int_{\mathcal X} dx \,\pTE(x;\bm\theta) \Bigl[- \epsilon(x) + \bm\theta \cdot \bm B(x) + \F(\bm\theta)  
    \notag \\[-0.3\baselineskip]
    & \quad \hspace{7.3em} + \epsilon(x) - \bm \theta' \cdot \bm B(x) - \F(\bm\theta') \Bigr]  
    \notag \\[0.15\baselineskip]
    & \quad = \F(\bm\theta) - \F(\bm\theta') + 
    (\bm\theta-\bm\theta') \cdot \langle \bm B \rangle_{\pTE(\bm\theta)}.
    \label{pr_KL_F_gen}    
\end{align}
Substitutions $\bm\theta \mapsto \bm 0$ and $\bm\theta' \mapsto \bm\theta$ give an expression of $\Delta \mathcal{F} (\bm{\theta})$,
\begin{equation}
    \Delta\F(\bm\theta) = - \kB T \, \DKL [p_\eq\|\pTE(\bm\theta)] - \bm\theta \cdot \langle \bm B \rangle_{p_\eq},
    \label{pr_KL_F}
\end{equation}
where we used $\pTE(\bm 0)=p_\eq$. Note that a similar expression has been found in Ref.~\cite{gorban2010entropy}.

\subsection{Explicit expression of the minimum EP}
\label{subsec:apdx_lowerbound}

We prove Eq.~\eqref{eq:pythagoras}, which is one of the two elements of the proof of Eq.~\eqref{eq:Legendre_main}. Let $\bm\eta$ be any set of expectation values and $q$ be any distribution that satisfies $\langle \bm B \rangle_{q} = \bm\eta$. The given set of expectation values $\bm\eta$ determines the set of parameter values $\bm\theta(\bm\eta)$ and the tilted equilibrium $\pTE(\bm\theta(\bm\eta))$, which we write as $\bm\theta$ and $\pTE$ for conciseness. Equation~\eqref{eq:def_theta_eta} reads $\langle \bm B \rangle_{\pTE} = \bm\eta$ in this shorthand notation.
Then the three distributions, $q$, $\pTE$, and $p_\eq$, satisfy the \textit{generalized Pythagorean theorem}~\cite{amari2016information}:
\begin{equation}
    \DKL[q \| p_\eq]  = \DKL[q \| \pTE] + \DKL[\pTE \| p_\eq] .
    \label{pr_gen_pytha}
\end{equation}
To prove this relation, we calculate the difference as follows:
\begin{align}
    &\kB T\left\{ \DKL[q\|\pTE]+\DKL[\pTE\|p_\eq]-\DKL[q\|p_\eq]\right\} 
    \notag \\
    &=\kB T\int_{\mathcal X}dx\,[\pTE(x)-q(x)]\ln\frac{\pTE(x)}{p_\eq(x)}
    \notag \\
    &= \int_{\mathcal X}dx\,[\pTE(x)-q(x)] 
    \notag \\[-0.3\baselineskip]
    &\hspace{3.5em} \times \Bigl[- \epsilon(x)
    + \bm{\theta}\cdot\bm{B}(x) + \F(\bm{\theta})
    + \epsilon(x) - \F(\bm{0}) \Bigr]
    \notag\\
    &=\bm\theta \cdot \int_{\mathcal X}dx\,[\pTE(x)-q(x)] \bm{B}(x) 
    \notag \\
    &=0. 
\end{align}
The third equality follows from $\int_{\mathcal X} dx\,\pTE(x) = \int_{\mathcal X} dx\,q(x) = 1$, and the last equality is due to the relation of expectation values $\langle \bm B\rangle_{\pTE} = \bm\eta = \langle\bm B \rangle_q$. Using Eq.~\eqref{pr_KL_H}, we can rewrite Eq.~\eqref{pr_gen_pytha} as
\begin{equation}
    \Delta \H[q] = \kB T \DKL [ q \| \pTE] + \Delta \H [\pTE].
    \label{pr_H_pythagoras}
\end{equation}
Recalling the abbreviation $\pTE \equiv \pTE(\bm\theta(\bm\eta))$, Eq.~\eqref{pr_H_pythagoras} is identical to the desired relation in Eq.~\eqref{eq:pythagoras}. 

We note that inserting Eqs.~\eqref{eq:minimum_cost_closed} and \eqref{eq:Legendre_math_b} into Eq.~\eqref{pr_H_pythagoras} gives an expression like the Donsker--Varadahn representation~\cite{donsker1983asymptotic}, but the use of coordinates $\bm\theta$ and the restriction of the set of observables are unique to information geometry.

\subsection{Legendre duality}
\label{subsec:apdx_Legendre}
We prove the Legendre transform between $\Delta \Hm(\bm\eta)$ and $\Delta  \F(\bm\theta)$ in Eq.~\eqref{eq:Legendre_math}, which is the second element of the proof of Eq.~\eqref{eq:Legendre_main}. 

First, we prove the Legendre transform from $\Delta \F(\bm\theta)$ to $\Delta \Hm(\bm\eta)$. We fix an arbitrary set of parameters $\bm\theta$ and use $\bm\eta \equiv \bm\eta(\bm\theta) = \langle \bm B\rangle_{\pTE(\bm\theta)} $ for the corresponding set of expectation values. The derivative relation $\partial \Delta \F /\partial \theta_\alpha = -\eta_\alpha$ is proved as
\begin{align}
    \frac{\partial \Delta \F}{\partial \theta_\alpha }
    &= - \kB T \frac{\partial}{\partial \theta_\alpha } \ln \qty[ {\int_{\mathcal X} dx} \exp\qty(-\frac{\epsilon(x)- \bm\theta \cdot \bm B(x)}{k_\mathrm B T}) ] 
    \notag \\
    &=- \frac{ \int_{\mathcal X} dx \, B_\alpha(x) \exp\{ - [\epsilon(x)- \bm\theta \cdot \bm B(x)] / k_\mathrm B T \} }{\int_{\mathcal X} dx\, \exp \{  -[\epsilon(x) -  \bm\theta \cdot \bm B(x)] / {k_\mathrm B T} \} }
    \notag \\
    &= -\int_{\mathcal X} dx \, B_\alpha(x) \pTE(x;\bm\theta) 
    \notag \\
    &= -\eta_\alpha.
    \label{pr_Legendre_derivative}
\end{align}
This relation states that the derivative of the equilibrium free energy by an external field parameter gives the corresponding expectation value, which is a classical result in statistical mechanics (e.g., Ref.~\cite{ruelle1969}). To prove the relation between the two functions, $\Delta \Hm(\bm\eta) = \Delta \F(\bm\theta) + \sum_\alpha \theta_\alpha\eta_\alpha$, we use the expression of $\Delta\Hm(\bm\eta)$ in Eq.~\eqref{eq:minimum_cost_closed} and find
\begin{align}
    \Delta \Hm(\bm\eta)
    &= \Delta \H[\pTE(\bm\theta)] 
    \notag \\
    &= \kB T\DKL[\pTE(\bm\theta)\|p_\eq] 
    \notag \\
    &= \F(\bm\theta)  - \F(\bm 0) + (\bm\theta - \bm 0) \cdot \langle \bm B\rangle_{\pTE(\bm\theta)}
    \notag \\
    &= \Delta\F(\bm\theta) + \bm\theta \cdot \bm\eta ,
\end{align}
where we have used Eqs.~\eqref{pr_KL_H} and \eqref{pr_KL_F_gen}.

Next, we prove the inverse transformation from $\Delta\Hm(\bm\eta)$ to $\Delta\F(\bm\theta)$. For this purpose, it suffices to show that $\Delta\F(\bm\theta)$ is a strictly concave function. Then the general theory of Legendre duality for strictly convex or strictly concave functions (e.g., Ref.~\cite{callen1985}) leads to the inverse transform $\partial \Delta \Hm / \partial \eta_\alpha = \theta_\alpha $ and $\Delta \F(\bm\theta) =\Delta  \Hm(\bm\eta) - \sum_\alpha \theta_\alpha\eta_\alpha$. The general theory also ensures that $\Delta\Hm$ is strictly convex.

To prove the strict concavity of $\Delta \F$, we consider two sets of parameter values $\bm\theta,\bm\theta'$. We consider the graph of the function $\Delta\F(\,\cdot\,)$ and its tangent plane at $\bm\theta$. The tangent plane, which we write as $\mathcal T(\,\cdot\,;\bm\theta)$, is given by
\begin{equation}
    \mathcal T(\bm\theta'; \bm\theta) = \Delta \F(\bm\theta) + \sum_\alpha \frac{\partial \Delta \mathcal F(\bm\theta)}{\partial \theta_\alpha} (\theta'_\alpha - \theta_\alpha).
\end{equation}
This tangent plane is always above the graph of $\F(\,\cdot\,)$:
\begin{align}
    &\mathcal T(\bm\theta';\bm\theta) - \Delta\F(\bm\theta') 
    \notag \\
    &\quad  =  \Delta\F(\bm\theta) - \Delta\F(\bm\theta') -  \langle \bm B \rangle_{\pTE(\bm\theta)} \cdot (\bm\theta' - \bm\theta) 
    \notag \\ 
    &\quad  = \kB T \DKL[ \pTE(\bm\theta) \| \pTE(\bm\theta')]
    \notag \\
    &\quad \geq 0,
\end{align}
where we have used the derivative relation in Eq.~\eqref{pr_Legendre_derivative} in the first equality, and the second equality follows from Eq.~\eqref{pr_KL_F_gen}. The last inequality holds with equality if and only if $\pTE(\bm\theta')=\pTE(\bm\theta)$. This happens if and only if  $\bm\theta' = \bm\theta$ since we have assumed that the observables $\bm B$ and the constant observable are linearly independent. Thus, the function $\Delta\F$ is strictly concave.

\subsection{Thermodynamic force}
\label{subsec:apdx_thermo_force}

We derive the relation between the thermodynamic force and the external fields in Eq.~\eqref{eq:thermo_force} under the realizability condition. Since the realizability condition implies $\pns(t) = \pTE(\bm\theta(\bmetans(t)))$ (Sec.~\ref{subsec:scaling}), we have
\begin{align}
    & \ln\frac{\pns(x;t)}{\pns(y;t)} 
    =  \ln\frac{\pTE(x;\bm\theta(\bmetans(t)))}{\pTE(y;\bm\theta(\bmetans(t)))} 
    \notag \\
    & \qquad =   - \frac{\epsilon(x) - \bm\theta(\bmetans(t)) \cdot \bm B(x)}{\kB T}  + \frac{\epsilon(y) - \bm\theta(\bmetans(t)) \cdot \bm B(y)}{\kB T}
    \notag \\
    &\qquad  = - \frac{\epsilon(x) - \epsilon(y)}{\kB T} + \frac{ \bm \theta(\bmetans (t)) \cdot [\bm B(x)- \bm B(y)]}{\kB T}.
\end{align}
Comparing this relation with the definition of the thermodynamic force in Eq.~\eqref{eq:thermo_force_def}, it is easy to prove Eq.~\eqref{eq:thermo_force}. The continuous-state version is proved similarly by replacing the differences in quantities between two states by their spatial derivatives, e.g., $\epsilon(x) - \epsilon(y)$ by $\nabla \epsilon(x)$.

\subsection{Monotonicity}
\label{subsec:apdx_monotonicity}
We prove the monotonicity of the function $\mathcal L(\bm\theta)$ in Eq.~\eqref{eq:monotonicity}, assuming the realizability condition and a Markovian time evolution. Comparing Eqs.~\eqref{pr_KL_F} and \eqref{eq:monotonicity_function}, we obtain $\mathcal L(\bm\theta) = -\kB T \DKL [p_\eq \| \pTE (\bm\theta)] $ for any $\bm\theta$. Combined with the realizability condition, we get
\begin{equation}
    \frac{\mathcal L(\bm\theta(\bmetans(t)))}{\kB T} = - \DKL[p_\eq \| \pTE(\bm\theta(\bmetans(t)))] = -\DKL [p_\eq \| \pns(t)].
    \label{pr_L_KL}
\end{equation}
For Markovian time evolutions, the KL divergence has the contraction property~\cite{kampen2007stochastic},
\begin{equation}
    d_t  \DKL[\tilde q(t) \| \pns(t)]  \leq 0,
    \label{eq:contraction}
\end{equation}
for two arbitrary trajectories, $\pns(t)$ and $\tilde q(t)$, obeying the same Markovian time evolution. By substituting $p_\eq$ for $\tilde q(t)$ and noting that $p_\eq$ is the fixed point of the time evolution, we obtain $d_t  \DKL[p_\eq \| \pns(t) ] \leq 0$. Combining this equation with Eq.~\eqref{pr_L_KL} proves the desired monotonicity in Eq.~\eqref{eq:monotonicity}.

Note that substituting $p_\eq$ for $\pns (t)$ in Eq.~\eqref{eq:contraction} and using the expression in Eq.~\eqref{pr_KL_H} gives the monotonicity $d_t \Delta\H[\pns(t)] \leq 0$, which is the second law of thermodynamics~\cite{vandenbroeck2015ensemble}. Therefore, the monotonicities of $\mathcal L$ and $\Delta\H$ have similar mathematical origins.

\section{Perturbative calculation of the error}
\label{sec:apdx_perturbation}

We compare a reference system that satisfies the realizability condition with a slightly different (perturbed) system that does not fulfill the realizability condition. Our goal is to show that the discrepancy between $\pns (t)$  and $\pTE (\bm\theta(\bmetans(t)))$ scales linearly with the magnitude of the perturbation.

We specify the reference system by the Markovian time evolution generator $\mathscr{L}$, the state energy $\epsilon \equiv \{\epsilon(x)\}_{x\in\mathcal X}$, the initial distribution $\pns(0) \equiv \{\pns(x;0)\}_{x\in\mathcal X}$, and the set of observables $\bm B \equiv \{\bm{B}(x)\}_{x\in\mathcal X}$. The Markovian time evolution generator, used only in this appendix, generates the time evolution as $d_{t}\pns(t)=\mathscr{L}\pns(t)$. We do not assume any particular form of it, only imposing the relaxation to the equilibrium $p_\eq(x)\propto e^{-\epsilon(x)}$. Here and until the end of this appendix, we set $\kB T=1$ for simplicity.

At a fixed time $t$, the reference system has the distribution $\pns(t)$, the set of expectation values $\bmetans(t) = \langle \bm B \rangle_{\pns(t)}$, the corresponding set of external field parameters $\bm\theta(\bmetans(t))$, and the tilted equilibrium $\pTE(\bm\theta(\bmetans(t)))$. For conciseness, we use the symbols $\pns$, $\bmetans$, $\bm\theta$, and $\pTE$ to denote these quantities for the reference system at the fixed time $t$. 
We assume that the reference system satisfies the realizability condition. As discussed in Sec.~\ref{subsec:scaling}, the realizability condition is equivalent to $\pns(t)=\pTE(\bm{\theta}(\bmetans(t)))$, which reads $\pns = \pTE$ in the shorthand notation adopted here.

We consider a small parameter $\lambda$ ($\vert\lambda\vert\ll1$) and a perturbed system with a generator $\mathscr{L}+\lambda\mathscr{L}' + o(\lambda)$, a state energy $\epsilon+\lambda\epsilon' + o(\lambda)$, an initial distribution $\pns(0)+\lambda\pns'(0) + o(\lambda)$, and a set of observables $\bm{B}+\lambda\bm{B}' + o(\lambda)$. Some but not all of $\gen'$, $\epsilon'$, $\tilde{p}'(0)$, and $\bm{B}'$ may be zero. These perturbations make small changes to $\pns(t)$, $\bmetans(t)$, $\bm{\theta}(\bmetans(t))$, and $\pTE(\bm{\theta}(\bmetans(t)))$ at the fixed time $t$. We write these quantities of the perturbed system as $\pns + \lambda^\sca \pns' + o(\lambda^\sca)$, $\bmetans + \lambda^\scb \bmetans' + o(\lambda^\scb)$, $\bm\theta + \lambda^\scc \bm\theta' + o(\lambda^\scc)$, and $\pTE + \lambda^\scd \pTE' + o(\lambda^\scd)$, where $\sca$, $\scb$, $\scc$, and $\scd$ are unknown positive exponents. In the following, we calculate these quantities and determine the exponents.

First, the perturbed probability distribution at time $t$ is 
\begin{align}
    & \pns + \lambda^\sca \pns'
    \notag \\
    & \simeq e^{(\gen+\lambda\gen')t}[\tilde{p}(0)+\lambda\tilde{p}'(0)]
    \notag \\
    & \simeq e^{\gen t}\pns(0) 
    + \lambda e^{\gen t}\pns'(0)
    + \lambda \int_{0}^{t}ds\,e^{\gen(t-s)}\gen'e^{\gen s} \pns(0),
    \label{pe_distribution}
\end{align}
where $\simeq$ denotes equality to the leading order in $\lambda$. Since $\pns = e^{\gen t} \pns(0)$ for the reference system, the perturbation term has the exponent $\sca = 1$. Note that we cannot exclude the possibility that the $O(\lambda)$ term in Eq.~\eqref{pe_distribution} vanishes identically, but in that case, we can still set $\beta=1$ and write Eq.~\eqref{pe_distribution} as $\pns + \lambda \pns' + o(\lambda)$ with $\pns'=0$. This caveat also applies to $\lambda^\scb \bmetans'$, $\lambda^\scc \bm\theta'$, and $\lambda^\scd \pTE'$ below.

The perturbed expectation values of the observables are 
\begin{align}
    &\bmetans + \lambda^\scb \bmetans'
    \notag \\
    &\simeq  \int_{\mathcal X}dx\,[\pns(x)+\lambda\pns'(x)][\bm{B}(x)+\lambda\bm{B}'(x)]
    \notag \\
    & \simeq \int_{\mathcal X} dx\, \bigl[ \pns(x)\bm B(x) + \lambda \pns'(x) \bm B(x) + \lambda \pns(x) \bm B'(x) \bigr].
\end{align}
The reference system has the set of expectation values $\bmetans = \int_{\mathcal X}dx\,\pns(x) \bm B(x)$, and thus, the exponent of the perturbation term is $\scb = 1$.

Next, we evaluate the perturbations to the set of parameter values $\bm{\theta}$. It is determined by solving
\begin{equation}
    \int_{\mathcal{X}}dx\,[\pTE(x)+\lambda^{\scd}\pTE'(x)]\,[\bm{B}(x)+\lambda\bm{B}'(x)]\simeq\bmetans+\lambda\bmetans'
    \label{pe_pTE_solve-1}
\end{equation}
with respect to $\bm\theta + \lambda^\scc\bm\theta'$, where $\bm\theta + \lambda^\scc \bm\theta'$ is related to $\pTE + \lambda^\scd\pTE'$ as 
\begin{align}
     & \pTE(x)+\lambda^{\scd}\pTE'(x)
     \notag \\
     & \simeq\frac{e^{-\left\{ [\epsilon(x)+\lambda\epsilon'(x)]-(\bm{\theta}+\lambda^{\scc}\bm{\theta}')\cdot[\bm{B}(x)+\lambda\bm{B}'(x)]\right\} }}{\int_{\mathcal{X}}dx\,e^{-\left\{ [\epsilon(x)+\lambda\epsilon'(x)]-(\bm{\theta}+\lambda^{\scc}\bm{\theta}')\cdot[\bm{B}(x)+\lambda\bm{B}'(x)]\right\} }}
     \notag \\
     & \simeq\frac{\pTE(x)\left\{ 1-\lambda\epsilon'(x)+\lambda\bm{\theta}\cdot\bm{B}'(x)+\lambda^{\scc}\bm{\theta}'\cdot\bm{B}(x)\right\} }{\int_{\mathcal{X}}dx\,\pTE(x)\left\{ 1-\lambda\epsilon'(x)+\lambda\bm{\theta}\cdot\bm{B}'(x)+\lambda^{\scc}\bm{\theta}'\cdot\bm{B}(x)\right\} }
     \notag \\
     & = \frac{\pTE(x)\left\{ 1-\lambda\epsilon'(x)+\lambda\bm{\theta}\cdot\bm{B}'(x)+\lambda^{\scc}\bm{\theta}'\cdot\bm{B}(x)\right\} }{1-\lambda\langle\epsilon'\rangle+\lambda\bm{\theta}\cdot\langle\bm{B}'\rangle+\lambda^{\scc}\bm{\theta}'\cdot\langle\bm{B}\rangle}
     \notag \\
     & \simeq\pTE(x)\Bigl\{1 + \lambda\bigl[ -\epsilon'(x)  + \langle\epsilon'\rangle + \bm{\theta}\cdot\bm{B}'(x) - \bm{\theta}\cdot\langle\bm{B}'\rangle\bigr]
     \notag \\[-0.2\baselineskip]
     & \hspace{4em}+\lambda^{\scc}\bigl[\bm{\theta}'\cdot\bm{B}(x) - \bm{\theta}'\cdot\langle\bm{B}\rangle ]\Bigr\}.
     \label{pe_pTE_ansatz}
\end{align}
Here,  we use $\pTE(x)\propto e^{-[\epsilon(x)-\bm{\theta}\cdot\bm{B}(x)]}$ in the second equality and use $\langle\,\cdot\,\rangle$ as shorthand for $\langle\,\cdot\,\rangle_{\pTE}$ until the end of this paragraph. Inserting Eq.~\eqref{pe_pTE_ansatz} into the left-hand side of Eq.~\eqref{pe_pTE_solve-1}, we get
\begin{align}
     & \mbox{[left-hand side of Eq.~\eqref{pe_pTE_solve-1}]}
     \notag \\
     & \simeq \int_{\mathcal{X}}dx \, \Bigl\{\pTE(x)\bm{B}(x) + \lambda \pTE(x)\bm{B}'(x) \notag \\
     & \hspace{2em} + \lambda\pTE(x)\bigl[-\epsilon'(x) + \langle\epsilon'\rangle + \bm{\theta}\cdot\bm{B}'(x)  -\bm{\theta}\cdot\langle\bm{B}'\rangle\bigr]\bm{B}(x)\notag \\
     & \hspace{2em}+\lambda^{\scc}\pTE(x) \bigl[\bm{\theta}'\cdot\bm{B}(x) - \bm{\theta}'\cdot\langle\bm{B}\rangle\bigr] \bm{B}(x)\Bigr\}\notag \\
     & \simeq\langle\bm{B}\rangle+\lambda\left\{ \langle\bm{B}'\rangle-\llangle\epsilon',\bm{B}\rrangle+\llangle\bm{B},\bm{B}'{}^{\trp}\rrangle\bm{\theta}\right\} +\lambda^{\scc}\llangle\bm{B},\bm{B}^{\trp}\rrangle\bm{\theta}',
\end{align}
where we have defined the covariance $\llangle X,Y\rrangle\coloneqq\langle XY\rangle-\langle X\rangle\langle Y\rangle$. Inserting this into Eq.~\eqref{pe_pTE_solve-1}, the resulting zeroth-order equation, $\langle\bm{B}\rangle=\bmetans$, is the equation for determining $\bm\theta$ of the reference system, and therefore, it is already satisfied. The remaining terms give the equation to determine $\lambda^{\scc}\bm{\theta}'$:
\begin{equation}
    \lambda^{\scc} \llangle\bm{B},\bm{B}^{\trp}\rrangle \bm{\theta}'
    \simeq \lambda\bmetans' - \lambda\left\{ \langle\bm{B}'\rangle - \llangle\epsilon',\bm{B}\rrangle+\llangle\bm{B},\bm{B}'{}^{\trp}\rrangle\bm{\theta}\right\} .
\end{equation}
Assuming that the covariance matrix $\llangle\bm{B},\bm{B}^{\trp}\rrangle$ is full rank, we obtain
\begin{equation}
    \lambda^{\scc}\bm{\theta}' 
    \simeq \lambda \left[ \llangle\bm{B},\bm{B}^{\trp}\rrangle \right]^{-1}\left\{ \bmetans'-\langle\bm{B}'\rangle+\llangle\epsilon',\bm{B}\rrangle  -\llangle \bm B, \bm B' {}^\trp \rrangle \bm\theta \right\} .
\end{equation}
This gives the exponent $\scc=1$.

Finally, we calculate the perturbation to $\pTE$.
Substituting $\scc=1$ back into Eq.~\eqref{pe_pTE_ansatz}, we find that the exponent of the perturbation to $\pTE$ is $\nu=1$.

Combining the above results, we can evaluate the discrepancy between the nonstationary distribution and the tilted equilibrium distribution for the perturbed system as 
\begin{align}
    & [\pns + \lambda \pns' + o(\lambda)]
    - [\pTE + \lambda \pTE' + o(\lambda)] 
    \notag \\*
    & \hspace{9em} = \lambda (\pns' - \pTE') + o(\lambda),
    \label{pe_conclusion}
\end{align}
where we have used the realizability condition for the reference system. From Eq.~\eqref{pe_conclusion}, it is fair to say that the discrepancy generically scales linearly in $\lambda$, which is the fact we used in Sec.~\ref{subsec:scaling} to evaluate the scaling of the error in the inference of the true EP\@. The exception is when $\pns' - \pTE'$ happens to vanish, in which case the scaling of the discrepancy is of higher order, and the error is even smaller.

\section{Generalizations of the setup}
\label{sec:apdx_generalized_setup}

We discuss two generalizations of the setup. Appendix~\ref{subsec:apdx_generalized_setup_chempot} deals with the inclusion of internal entropy and a particle reservoir. Appendix~\ref{subsec:apdx_generalized_setup_NESS} considers a formal extension for systems with multiple baths.

\subsection{Systems with internal entropy and a particle reservoir}
\label{subsec:apdx_generalized_setup_chempot}

We can generalize all our results to systems with internal entropy and/or in contact with a single particle reservoir. The internal entropy $s_\inter(x)$ is an additional contribution to the system entropy when the system is in state $x$. It enters the system entropy as $ \int_{\mathcal X} dx\, p(x) \qty[ s_\inter(x) - \kB\ln p(x) ]$. Such a contribution arises, for example, when the states are already coarse-grained~\cite{esposito2012stochastic}. We also allow the system to exchange particles with a single particle reservoir of chemical potential $\mu$. The particle number of the system in state $x$ is denoted by $n(x)$.

These generalizations modify the definitions of EP in Eq.~\eqref{eq:noneqH} and tilted equilibrium free energy in Eq.~\eqref{eq:tilted_F} as~\cite{esposito2012stochastic}
\begin{subequations}
\label{ge_int_res}
\begin{align}
    \H[p] &= \kB T \!\int_{\mathcal X} dx\, p(x) \ln p(x) + \int_{\mathcal X} dx\, \omega(x) p(x), \\
    \F(\bm\theta) &= - \kB T \ln \qty[ \int_{\mathcal X} dx \exp \qty(-\frac{\omega(x) - v(x;\bm\theta)}{\kB T}) ],
\end{align}
\end{subequations}
where $\omega(x) \coloneqq \epsilon(x) - \mu n(x) - Ts_\inter(x)$. This definition of $\H[p]$ is equal to the sum of $(-T)$ times the system entropy, $ \int_{\mathcal X} dx\, p(x) [-Ts_\inter(x) + \kB T\ln p(x) ]$, and $(-T)$ times the bath entropy, $\int_{\mathcal X} dx\, p(x)\qty[\epsilon(x) - \mu n(x)]$, where we neglected a constant term in the bath entropy that does not depend on $p$. Therefore, the difference $\Delta\H[p] = \H[p] - \H[p_\eq]$ correctly gives the EP of the relaxation process from $p$ to $p_\eq$. The equilibrium distribution of this system is $p_\eq(x) \propto \exp \qty[ -\omega(x)/ \kB T]$~\cite{esposito2012stochastic}.

Since the modified definitions in Eq.~\eqref{ge_int_res} are formally obtained by replacing $\epsilon(x)$ with $\omega(x)$ in the original definitions, all our results can be generalized with this replacement. Since $\epsilon(x)$ does not appear explicitly in our inference methods, we can collect the nonstationary data and the tilted equilibrium data by the same procedure as in the main text, and we can calculate thermodynamic quantities from the data without modifying the equations. An external field $v(x;\bm\theta)$ physically changes the energy from $\epsilon(x)$ to $\epsilon(x) - v(x;\bm\theta)$, but it can also be interpreted as changing $\omega(x)$ to $\omega(x) - v(x;\bm\theta)$, as follows from the definition of $\omega(x)$.

\subsection{Systems coupled with multiple baths}
\label{subsec:apdx_generalized_setup_NESS}

We can also formally generalize our results to systems in contact with multiple heat baths and/or multiple particle reservoirs. Consider a system that relaxes to a nonequilibrium stationary distribution $p_\st \equiv \{p_\st(x)\}_{x\in\mathcal X}$ incurring a nonzero EP rate.
To generalize our results for such systems, we define $\phi(x)$ by $p_\st(x) = \exp [-\phi(x)]$~\cite{hatano2001steady}, and we replace $\epsilon(x)/\kB T$ with $\phi(x)$ and $p_\eq(x)$ with $p_\st(x)$. We introduce $\hat\H$, $\hat\F$, and $\hat v$ to replace $\H/\kB T$, $\F/\kB T$, and $v/\kB T$, respectively, to eliminate $T$ from the theory because the temperature $T$ is not defined when the system is in contact with multiple heat baths. With these replacements, we can formally redefine the EP and the tilted equilibrium free energy as
\begin{subequations}
\label{ge_nonDB}
\begin{align}
    \hat\H[p] &= \int_{\mathcal X} dx\, p(x) \ln p(x) +  \int_{\mathcal X} dx\, \phi(x) p(x), \\
    \hat\F(\bm\theta) &= -  \ln \qty[ \int_{\mathcal X} dx \exp \qty{-[\phi(x) - \hat v(x;\bm\theta)] \strut} ].
\end{align}
\end{subequations}
The difference $\Delta\hat\H[p] = \hat\H[p] - \hat\H[p_\st]$ is known as the Hatano--Sasa excess EP~\cite{hatano2001steady} and the nonadiabatic EP~\cite{esposito2010three} over the process from $p$ to $p_\st$.

We can formally reproduce all our results with these replacements. If we can physically implement an external field that modifies the stationary distribution $p_\st$ to a tilted stationary distribution proportional to $\exp\{- [\phi(x) - \hat v(x;\bm\theta) ] \}$, we can use the procedure in the main text to obtain the minimum of $\Delta\hat\H[p]$ compatible with the observed expectation values, and we can also calculate other thermodynamic quantities under the realizability condition. However, this generalization remains formal because we cannot physically realize the tilted stationary distribution generically.

\section{Supplemental information on the example}
\label{sec:apdx_example}

\subsection{Details of the numerical calculations}
\label{subsec:apdx_numerics}

In the numerical calculations (Figs.~\ref{fig:example}--\ref{fig:example_additional}), we numerically solved the Fokker--Planck equation by discretizing the space and solving the resulting discrete-space continuous-time Markov jump system. We calculated the expectation values of the observables, the EP, and the tilted equilibrium free energy by replacing the integrals with the sums over the discretized space. We confirmed that the results do not depend on the spatial mesh size. We also checked that the results are consistent with the analytical calculation when $\epsilon_r(x) = 0$, which is detailed in the next section.

\subsection{Analytical calculation for the harmonic system}
\label{subsec:apdx_example_analytic}

We analytically calculate the relevant quantities for the example system in the main text (Sec.~\ref{sec:example}) with $\epsilon(x)=ax^2$, i.e., in the absence of the ragged potential. 
Assuming a Gaussian initial distribution, the system satisfies the realizability condition, as discussed in the main text.

The time evolution is governed by the Fokker--Planck equation in Sec.~\ref{subsec:example_setup} of the main text. The nonstationary distribution $\pns(t)$ is the Gaussian distribution with mean $\tilde c_1(t)$ and variance $\tilde c_2(t)$ with
\begin{subequations}
\begin{align}
    \tilde c_1(t) &= \tilde c_1(0) e^{-2\mu a t},\\
    \tilde c_2(t) &= \qty[ \tilde c_2(0) - \frac{\kB T}{2a} ] e^{-4\mu a t} + \frac{\kB T}{2a}.
\end{align}
\end{subequations}
In the nonstationary measurement, we measure the expectation values of the observables, $B_1(x)=x$ and $B_2(x)=x^2$. The expectation values are explicitly given by $\tilde\eta_1(t) = \langle x \rangle_{\pns(t)} =  \tilde c_1(t)$ and $\tilde\eta_2(t) = \langle x^2 \rangle_{\pns(t)} = \tilde c_2(t) + [\tilde c_1(t)]^2$.

In the tilted equilibrium measurements, we apply the external fields in Eq.~\eqref{eq:ext_field} and measure the expectation values of the observables $B_1$ and $B_2$. The modified state energy due to the external fields is
\begin{align}
    &\epsilon(x) - v(x;\bm\theta) = a x^2 - \theta_1 x - \theta_2 x^2 
    \notag \\
    &\quad =  \qty(a - \theta_2) \qty(x - \frac{\theta_1}{2\qty(a - \theta_2)})^2 - \frac{(\theta_1)^2}{4\qty( a - \theta_2)}.
    \label{ex_e-v}
\end{align}
The modified state energy is harmonic, and therefore, the tilted equilibrium is a Gaussian distribution with mean $\theta_1 /2(a-\theta_2)$ and variance $\kB T / 2(a-\theta_2)$. The set of expectation values at the tilted equilibrium is
\begin{equation}
    \eta_1(\bm\theta) = \frac{\theta_1}{2(a-\theta_2)}, \quad
    \eta_2(\bm\theta) = \frac{\kB T}{2(a-\theta_2)} + \frac{(\theta_1)^2}{4(a-\theta_2)^2}.
    \label{ga_eta}
\end{equation} 
The tilted equilibrium free energy is
\begin{equation}
    \Delta\F (\bm\theta) 
    = \frac{\kB T}{2} \ln \qty(\frac{a - \theta_2}{a}) - \frac{(\theta_1)^2}{4\qty(a - \theta_2)}.
    \label{ex_FB}
\end{equation}
The inverse correspondence from $\bm\eta$ to $\bm\theta$ is 
\begin{equation}
    \theta_1(\bm\eta) = \frac{\kB T  \eta_1}{\eta_2 - (\eta_1)^2}, \quad 
    \theta_2(\bm\eta) = a - \frac{\kB T}{2[ \eta_2 - (\eta_1)^2]},
    \label{ex_theta_eta}
\end{equation}
where the term $\eta_2 - (\eta_1)^2$ is the variance of the tilted equilibrium distribution.

From these data, we can obtain the following properties of the relaxation process. First, $\Delta\Hm(\bm\eta)$  is calculated from the right-hand side of Eq.~\eqref{eq:Legendre_main}, which is explicitly given by
\begin{align}
    \Delta\Hm(\bm\eta) &= \Delta \F (\bm \theta(\bm\eta)) + \theta_1(\bm\eta) \eta_1 + \theta_2(\bm\eta) \eta_2 
    \notag \\*
    & = - \frac{\kB T }{2} \ln \qty(\frac{2a[\eta_2 - (\eta_1)^2]}{\kB T }) + a\eta_2 - \frac{\kB T}{2}.
    \label{ex_Legendre}
\end{align}
One can directly check that Eq.~\eqref{ex_Legendre} is equal to the EP $\Delta \H[p]$ for $\epsilon(x) = ax^2$ with $p$ being the Gaussian distribution with mean $\eta_1$ and variance $\eta_2 - (\eta_1)^2$. 
Second, we obtain a decomposition of the EP rate:
\begin{equation}
    \epr(t) = -\frac{1}{T} \bigl[ \theta_1(\bmetans(t))\, d_t \tilde\eta_1(t) + \theta_2(\bmetans(t))\, d_t \tilde\eta_2(t) \bigr],
\end{equation}
with $\bm\theta(\bm\eta)$ in Eq.~\eqref{ex_theta_eta}. 
Third, we obtain the thermodynamic force from the continuous version of Eq.~\eqref{eq:thermo_force} as
\begin{align}
    \tilde f(x;t)
    &= - \frac {1}{T} [ \theta_1(\bmetans(t)) + 2 \theta_2(\bmetans(t)) x ] 
    \notag \\
    &= \frac{1}{T} \left( \frac{\kB T[x - \tilde\eta_1(t)]}{\tilde\eta_2(t) - [\tilde\eta_1(t)]^2} - 2ax\right).
\end{align}
Finally, the function $\mathcal L(\bm\theta)$ in Eq.~\eqref{eq:monotonicity_function} is given by 
\newpage
\begin{equation}
    \mathcal L (\bm\theta) 
    = \frac{\kB T}{2}  \ln \qty(\frac{a - \theta_2}{a}) - \frac{(\theta_1)^2}{4\qty(a - \theta_2)} + \theta_2 \frac{\kB T}{2a}.
    \label{ex_L}
\end{equation}
The function $\mathcal L(\bm\theta(\bmetans(t)))$ increases monotonically with time.
\newpage
\vspace*{3.5\baselineskip}


\begin{thebibliography}{72}%
\makeatletter
\providecommand \@ifxundefined [1]{%
 \@ifx{#1\undefined}
}%
\providecommand \@ifnum [1]{%
 \ifnum #1\expandafter \@firstoftwo
 \else \expandafter \@secondoftwo
 \fi
}%
\providecommand \@ifx [1]{%
 \ifx #1\expandafter \@firstoftwo
 \else \expandafter \@secondoftwo
 \fi
}%
\providecommand \natexlab [1]{#1}%
\providecommand \enquote  [1]{``#1''}%
\providecommand \bibnamefont  [1]{#1}%
\providecommand \bibfnamefont [1]{#1}%
\providecommand \citenamefont [1]{#1}%
\providecommand \href@noop [0]{\@secondoftwo}%
\providecommand \href [0]{\begingroup \@sanitize@url \@href}%
\providecommand \@href[1]{\@@startlink{#1}\@@href}%
\providecommand \@@href[1]{\endgroup#1\@@endlink}%
\providecommand \@sanitize@url [0]{\catcode `\\12\catcode `\$12\catcode
  `\&12\catcode `\#12\catcode `\^12\catcode `\_12\catcode `\%12\relax}%
\providecommand \@@startlink[1]{}%
\providecommand \@@endlink[0]{}%
\providecommand \url  [0]{\begingroup\@sanitize@url \@url }%
\providecommand \@url [1]{\endgroup\@href {#1}{\urlprefix }}%
\providecommand \urlprefix  [0]{URL }%
\providecommand \Eprint [0]{\href }%
\providecommand \doibase [0]{https://doi.org/}%
\providecommand \selectlanguage [0]{\@gobble}%
\providecommand \bibinfo  [0]{\@secondoftwo}%
\providecommand \bibfield  [0]{\@secondoftwo}%
\providecommand \translation [1]{[#1]}%
\providecommand \BibitemOpen [0]{}%
\providecommand \bibitemStop [0]{}%
\providecommand \bibitemNoStop [0]{.\EOS\space}%
\providecommand \EOS [0]{\spacefactor3000\relax}%
\providecommand \BibitemShut  [1]{\csname bibitem#1\endcsname}%
\let\auto@bib@innerbib\@empty
\bibitem [{\citenamefont {Bialeck}(2012)}]{Bialek2012Biophysics}%
  \BibitemOpen
  \bibfield  {author} {\bibinfo {author} {\bibfnamefont {W.}~\bibnamefont
  {Bialeck}},\ }\href@noop {} {\emph {\bibinfo {title} {Biophysics: {S}earching
  for principles}}}\ (\bibinfo  {publisher} {Princeton University Press},\
  \bibinfo {address} {New Jersey},\ \bibinfo {year} {2012})\BibitemShut
  {NoStop}%
\bibitem [{\citenamefont {Bhalla}\ and\ \citenamefont
  {Iyengar}(1999)}]{Bhalla1999EmergentProperties}%
  \BibitemOpen
  \bibfield  {author} {\bibinfo {author} {\bibfnamefont {U.~S.}\ \bibnamefont
  {Bhalla}}\ and\ \bibinfo {author} {\bibfnamefont {R.}~\bibnamefont
  {Iyengar}},\ }\bibfield  {title} {\bibinfo {title} {Emergent properties of
  networks of biological signaling pathways},\ }\href
  {https://doi.org/10.1126/science.283.5400.381} {\bibfield  {journal}
  {\bibinfo  {journal} {Science}\ }\textbf {\bibinfo {volume} {283}},\ \bibinfo
  {pages} {381} (\bibinfo {year} {1999})}\BibitemShut {NoStop}%
\bibitem [{\citenamefont {Dayan}\ and\ \citenamefont
  {Abbott}(2001)}]{dayan2001TheoreticalNeuroscience}%
  \BibitemOpen
  \bibfield  {author} {\bibinfo {author} {\bibfnamefont {P.}~\bibnamefont
  {Dayan}}\ and\ \bibinfo {author} {\bibfnamefont {L.~F.}\ \bibnamefont
  {Abbott}},\ }\href@noop {} {\emph {\bibinfo {title} {Theoretical
  Neuroscience}}}\ (\bibinfo  {publisher} {The MIT Press},\ \bibinfo {address}
  {Cambridge},\ \bibinfo {year} {2001})\BibitemShut {NoStop}%
\bibitem [{\citenamefont {Wark}\ \emph {et~al.}(2007)\citenamefont {Wark},
  \citenamefont {Lundstrom},\ and\ \citenamefont
  {Fairhall}}]{Wark2007SensoryAdaptation}%
  \BibitemOpen
  \bibfield  {author} {\bibinfo {author} {\bibfnamefont {B.}~\bibnamefont
  {Wark}}, \bibinfo {author} {\bibfnamefont {B.~N.}\ \bibnamefont
  {Lundstrom}},\ and\ \bibinfo {author} {\bibfnamefont {A.}~\bibnamefont
  {Fairhall}},\ }\bibfield  {title} {\bibinfo {title} {Sensory adaptation},\
  }\href {https://doi.org/10.1016/j.conb.2007.07.001} {\bibfield  {journal}
  {\bibinfo  {journal} {Curr. Opin. Neurobiol.}\ }\textbf {\bibinfo {volume}
  {17}},\ \bibinfo {pages} {423} (\bibinfo {year} {2007})}\BibitemShut
  {NoStop}%
\bibitem [{\citenamefont {Polettini}\ and\ \citenamefont
  {Esposito}(2013)}]{polettini2013nonconvexity}%
  \BibitemOpen
  \bibfield  {author} {\bibinfo {author} {\bibfnamefont {M.}~\bibnamefont
  {Polettini}}\ and\ \bibinfo {author} {\bibfnamefont {M.}~\bibnamefont
  {Esposito}},\ }\bibfield  {title} {\bibinfo {title} {Nonconvexity of the
  relative entropy for {M}arkov dynamics: A {F}isher information approach},\
  }\href {https://doi.org/10.1103/PhysRevE.88.012112} {\bibfield  {journal}
  {\bibinfo  {journal} {Phys. Rev. E}\ }\textbf {\bibinfo {volume} {88}},\
  \bibinfo {pages} {012112} (\bibinfo {year} {2013})}\BibitemShut {NoStop}%
\bibitem [{\citenamefont {Lu}\ and\ \citenamefont
  {Raz}(2017)}]{lu2017nonequilibrium}%
  \BibitemOpen
  \bibfield  {author} {\bibinfo {author} {\bibfnamefont {Z.}~\bibnamefont
  {Lu}}\ and\ \bibinfo {author} {\bibfnamefont {O.}~\bibnamefont {Raz}},\
  }\bibfield  {title} {\bibinfo {title} {Nonequilibrium thermodynamics of the
  {M}arkovian {M}pemba effect and its inverse},\ }\href
  {https://doi.org/10.1073/pnas.1701264114} {\bibfield  {journal} {\bibinfo
  {journal} {Proc. Natl. Acad. Sci.}\ }\textbf {\bibinfo {volume} {114}},\
  \bibinfo {pages} {5083} (\bibinfo {year} {2017})}\BibitemShut {NoStop}%
\bibitem [{\citenamefont {Kumar}\ and\ \citenamefont
  {Bechhoefer}(2020)}]{Kumar2020ExponentiallyFaster}%
  \BibitemOpen
  \bibfield  {author} {\bibinfo {author} {\bibfnamefont {A.}~\bibnamefont
  {Kumar}}\ and\ \bibinfo {author} {\bibfnamefont {J.}~\bibnamefont
  {Bechhoefer}},\ }\bibfield  {title} {\bibinfo {title} {Exponentially faster
  cooling in a colloidal system},\ }\href
  {https://doi.org/10.1038/s41586-020-2560-x} {\bibfield  {journal} {\bibinfo
  {journal} {Nature}\ }\textbf {\bibinfo {volume} {584}},\ \bibinfo {pages}
  {64} (\bibinfo {year} {2020})}\BibitemShut {NoStop}%
\bibitem [{\citenamefont {Berthier}\ and\ \citenamefont
  {Biroli}(2011)}]{Berthier2011TheoreticalPerspectiveOnTheGlass}%
  \BibitemOpen
  \bibfield  {author} {\bibinfo {author} {\bibfnamefont {L.}~\bibnamefont
  {Berthier}}\ and\ \bibinfo {author} {\bibfnamefont {G.}~\bibnamefont
  {Biroli}},\ }\bibfield  {title} {\bibinfo {title} {Theoretical perspective on
  the glass transition and amorphous materials},\ }\href
  {https://doi.org/10.1103/RevModPhys.83.587} {\bibfield  {journal} {\bibinfo
  {journal} {Rev. Mod. Phys.}\ }\textbf {\bibinfo {volume} {83}},\ \bibinfo
  {pages} {587} (\bibinfo {year} {2011})}\BibitemShut {NoStop}%
\bibitem [{\citenamefont {Diaconis}(1996)}]{Diaconis1996Cutoff}%
  \BibitemOpen
  \bibfield  {author} {\bibinfo {author} {\bibfnamefont {P.}~\bibnamefont
  {Diaconis}},\ }\bibfield  {title} {\bibinfo {title} {The cutoff phenomenon in
  finite {M}arkov chains.},\ }\href {https://doi.org/10.1073/pnas.93.4.1659}
  {\bibfield  {journal} {\bibinfo  {journal} {Proc. Natl. Acad. Sci.}\ }\textbf
  {\bibinfo {volume} {93}},\ \bibinfo {pages} {1659} (\bibinfo {year}
  {1996})}\BibitemShut {NoStop}%
\bibitem [{\citenamefont {Kastoryano}\ \emph {et~al.}(2012)\citenamefont
  {Kastoryano}, \citenamefont {Reeb},\ and\ \citenamefont
  {Wolf}}]{Kastoryano2012Cutoff}%
  \BibitemOpen
  \bibfield  {author} {\bibinfo {author} {\bibfnamefont {M.~J.}\ \bibnamefont
  {Kastoryano}}, \bibinfo {author} {\bibfnamefont {D.}~\bibnamefont {Reeb}},\
  and\ \bibinfo {author} {\bibfnamefont {M.~M.}\ \bibnamefont {Wolf}},\
  }\bibfield  {title} {\bibinfo {title} {A cutoff phenomenon for quantum
  {M}arkov chains},\ }\href {https://doi.org/10.1088/1751-8113/45/7/075307}
  {\bibfield  {journal} {\bibinfo  {journal} {J. Phys. A: Math. Theor.}\
  }\textbf {\bibinfo {volume} {45}},\ \bibinfo {pages} {075307} (\bibinfo
  {year} {2012})}\BibitemShut {NoStop}%
\bibitem [{\citenamefont {Schnakenberg}(1976)}]{schnakenberg1976network}%
  \BibitemOpen
  \bibfield  {author} {\bibinfo {author} {\bibfnamefont {J.}~\bibnamefont
  {Schnakenberg}},\ }\bibfield  {title} {\bibinfo {title} {Network theory of
  microscopic and macroscopic behavior of master equation systems},\ }\href
  {https://doi.org/10.1103/RevModPhys.48.571} {\bibfield  {journal} {\bibinfo
  {journal} {Rev. Mod. Phys.}\ }\textbf {\bibinfo {volume} {48}},\ \bibinfo
  {pages} {571} (\bibinfo {year} {1976})}\BibitemShut {NoStop}%
\bibitem [{\citenamefont {Seifert}(2012)}]{seifert2012stochastic}%
  \BibitemOpen
  \bibfield  {author} {\bibinfo {author} {\bibfnamefont {U.}~\bibnamefont
  {Seifert}},\ }\bibfield  {title} {\bibinfo {title} {Stochastic
  thermodynamics, fluctuation theorems and molecular machines},\ }\href
  {https://doi.org/10.1088/0034-4885/75/12/126001} {\bibfield  {journal}
  {\bibinfo  {journal} {Rep. Prog. Phys.}\ }\textbf {\bibinfo {volume} {75}},\
  \bibinfo {pages} {126001} (\bibinfo {year} {2012})}\BibitemShut {NoStop}%
\bibitem [{\citenamefont {Sekimoto}(2010)}]{sekimoto2010stochastic}%
  \BibitemOpen
  \bibfield  {author} {\bibinfo {author} {\bibfnamefont {K.}~\bibnamefont
  {Sekimoto}},\ }\href@noop {} {\emph {\bibinfo {title} {Stochastic
  Energetics}}}\ (\bibinfo  {publisher} {Springer},\ \bibinfo {address}
  {Berlin},\ \bibinfo {year} {2010})\BibitemShut {NoStop}%
\bibitem [{\citenamefont {Seifert}(2019)}]{seifert2019stochastic}%
  \BibitemOpen
  \bibfield  {author} {\bibinfo {author} {\bibfnamefont {U.}~\bibnamefont
  {Seifert}},\ }\bibfield  {title} {\bibinfo {title} {From stochastic
  thermodynamics to thermodynamic inference},\ }\href
  {https://doi.org/10.1146/annurev-conmatphys-031218-013554} {\bibfield
  {journal} {\bibinfo  {journal} {Annu. Rev. Condens. Matter Phys.}\ }\textbf
  {\bibinfo {volume} {10}},\ \bibinfo {pages} {171} (\bibinfo {year}
  {2019})}\BibitemShut {NoStop}%
\bibitem [{\citenamefont {Rold\'an}\ and\ \citenamefont
  {Parrondo}(2010)}]{roldan2010estimating}%
  \BibitemOpen
  \bibfield  {author} {\bibinfo {author} {\bibfnamefont {E.}~\bibnamefont
  {Rold\'an}}\ and\ \bibinfo {author} {\bibfnamefont {J.~M.~R.}\ \bibnamefont
  {Parrondo}},\ }\bibfield  {title} {\bibinfo {title} {Estimating dissipation
  from single stationary trajectories},\ }\href
  {https://doi.org/10.1103/PhysRevLett.105.150607} {\bibfield  {journal}
  {\bibinfo  {journal} {Phys. Rev. Lett.}\ }\textbf {\bibinfo {volume} {105}},\
  \bibinfo {pages} {150607} (\bibinfo {year} {2010})}\BibitemShut {NoStop}%
\bibitem [{\citenamefont {Rold\'an}\ and\ \citenamefont
  {Parrondo}(2012)}]{Roldan2012EntropyProduction}%
  \BibitemOpen
  \bibfield  {author} {\bibinfo {author} {\bibfnamefont {E.}~\bibnamefont
  {Rold\'an}}\ and\ \bibinfo {author} {\bibfnamefont {J.~M.~R.}\ \bibnamefont
  {Parrondo}},\ }\bibfield  {title} {\bibinfo {title} {Entropy production and
  {K}ullback-{L}eibler divergence between stationary trajectories of discrete
  systems},\ }\href {https://doi.org/10.1103/PhysRevE.85.031129} {\bibfield
  {journal} {\bibinfo  {journal} {Phys. Rev. E}\ }\textbf {\bibinfo {volume}
  {85}},\ \bibinfo {pages} {031129} (\bibinfo {year} {2012})}\BibitemShut
  {NoStop}%
\bibitem [{\citenamefont {Muy}\ \emph {et~al.}(2013)\citenamefont {Muy},
  \citenamefont {Kundu},\ and\ \citenamefont {Lacoste}}]{muy2013noninvasive}%
  \BibitemOpen
  \bibfield  {author} {\bibinfo {author} {\bibfnamefont {S.}~\bibnamefont
  {Muy}}, \bibinfo {author} {\bibfnamefont {A.}~\bibnamefont {Kundu}},\ and\
  \bibinfo {author} {\bibfnamefont {D.}~\bibnamefont {Lacoste}},\ }\bibfield
  {title} {\bibinfo {title} {Non-invasive estimation of dissipation from
  non-equilibrium fluctuations in chemical reactions},\ }\href
  {https://doi.org/10.1063/1.4821760} {\bibfield  {journal} {\bibinfo
  {journal} {J. Chem. Phys.}\ }\textbf {\bibinfo {volume} {139}},\ \bibinfo
  {pages} {09B645\_1} (\bibinfo {year} {2013})}\BibitemShut {NoStop}%
\bibitem [{\citenamefont {Harada}\ and\ \citenamefont
  {Sasa}(2005)}]{Harada2005PRL}%
  \BibitemOpen
  \bibfield  {author} {\bibinfo {author} {\bibfnamefont {T.}~\bibnamefont
  {Harada}}\ and\ \bibinfo {author} {\bibfnamefont {S.-i.}\ \bibnamefont
  {Sasa}},\ }\bibfield  {title} {\bibinfo {title} {Equality connecting energy
  dissipation with a violation of the fluctuation-response relation},\ }\href
  {https://doi.org/10.1103/PhysRevLett.95.130602} {\bibfield  {journal}
  {\bibinfo  {journal} {Phys. Rev. Lett.}\ }\textbf {\bibinfo {volume} {95}},\
  \bibinfo {pages} {130602} (\bibinfo {year} {2005})}\BibitemShut {NoStop}%
\bibitem [{\citenamefont {Harada}\ and\ \citenamefont
  {Sasa}(2006)}]{Harada2006PRE}%
  \BibitemOpen
  \bibfield  {author} {\bibinfo {author} {\bibfnamefont {T.}~\bibnamefont
  {Harada}}\ and\ \bibinfo {author} {\bibfnamefont {S.-i.}\ \bibnamefont
  {Sasa}},\ }\bibfield  {title} {\bibinfo {title} {Energy dissipation and
  violation of the fluctuation-response relation in nonequilibrium langevin
  systems},\ }\href {https://doi.org/10.1103/PhysRevE.73.026131} {\bibfield
  {journal} {\bibinfo  {journal} {Phys. Rev. E}\ }\textbf {\bibinfo {volume}
  {73}},\ \bibinfo {pages} {026131} (\bibinfo {year} {2006})}\BibitemShut
  {NoStop}%
\bibitem [{\citenamefont {Rahav}\ and\ \citenamefont
  {Jarzynski}(2007)}]{rahav2007fluctuation}%
  \BibitemOpen
  \bibfield  {author} {\bibinfo {author} {\bibfnamefont {S.}~\bibnamefont
  {Rahav}}\ and\ \bibinfo {author} {\bibfnamefont {C.}~\bibnamefont
  {Jarzynski}},\ }\bibfield  {title} {\bibinfo {title} {Fluctuation relations
  and coarse-graining},\ }\href
  {https://doi.org/10.1088/1742-5468/2007/09/P09012} {\bibfield  {journal}
  {\bibinfo  {journal} {J. Stat. Mech.: Theory Exp.}\ }\textbf {\bibinfo
  {volume} {2007}}\bibinfo  {number} { (09)},\ \bibinfo {pages}
  {P09012}}\BibitemShut {NoStop}%
\bibitem [{\citenamefont {Esposito}(2012)}]{esposito2012stochastic}%
  \BibitemOpen
\bibfield  {number} {  }\bibfield  {author} {\bibinfo {author} {\bibfnamefont
  {M.}~\bibnamefont {Esposito}},\ }\bibfield  {title} {\bibinfo {title}
  {Stochastic thermodynamics under coarse graining},\ }\href
  {https://doi.org/10.1103/PhysRevE.85.041125} {\bibfield  {journal} {\bibinfo
  {journal} {Phys. Rev. E}\ }\textbf {\bibinfo {volume} {85}},\ \bibinfo
  {pages} {041125} (\bibinfo {year} {2012})}\BibitemShut {NoStop}%
\bibitem [{\citenamefont {Bo}\ and\ \citenamefont
  {Celani}(2014)}]{Bo2014EntropyProduction}%
  \BibitemOpen
  \bibfield  {author} {\bibinfo {author} {\bibfnamefont {S.}~\bibnamefont
  {Bo}}\ and\ \bibinfo {author} {\bibfnamefont {A.}~\bibnamefont {Celani}},\
  }\bibfield  {title} {\bibinfo {title} {Entropy production in stochastic
  systems with fast and slow time-scales},\ }\href
  {https://doi.org/10.1007/s10955-014-0922-1} {\bibfield  {journal} {\bibinfo
  {journal} {J. Stat. Phys.}\ }\textbf {\bibinfo {volume} {154}},\ \bibinfo
  {pages} {1325} (\bibinfo {year} {2014})}\BibitemShut {NoStop}%
\bibitem [{\citenamefont {Skinner}\ and\ \citenamefont
  {Dunkel}(2021{\natexlab{a}})}]{Skinner2021Improved}%
  \BibitemOpen
  \bibfield  {author} {\bibinfo {author} {\bibfnamefont {D.~J.}\ \bibnamefont
  {Skinner}}\ and\ \bibinfo {author} {\bibfnamefont {J.}~\bibnamefont
  {Dunkel}},\ }\bibfield  {title} {\bibinfo {title} {Improved bounds on entropy
  production in living systems},\ }\href
  {https://doi.org/10.1073/pnas.2024300118} {\bibfield  {journal} {\bibinfo
  {journal} {Proc. Natl. Acad. Sci.}\ }\textbf {\bibinfo {volume} {118}},\
  \bibinfo {pages} {e2024300118} (\bibinfo {year}
  {2021}{\natexlab{a}})}\BibitemShut {NoStop}%
\bibitem [{\citenamefont {Shiraishi}\ and\ \citenamefont
  {Sagawa}(2015)}]{shiraishi2015fluctuation}%
  \BibitemOpen
  \bibfield  {author} {\bibinfo {author} {\bibfnamefont {N.}~\bibnamefont
  {Shiraishi}}\ and\ \bibinfo {author} {\bibfnamefont {T.}~\bibnamefont
  {Sagawa}},\ }\bibfield  {title} {\bibinfo {title} {Fluctuation theorem for
  partially masked nonequilibrium dynamics},\ }\href
  {https://doi.org/10.1103/PhysRevE.91.012130} {\bibfield  {journal} {\bibinfo
  {journal} {Phys. Rev. E}\ }\textbf {\bibinfo {volume} {91}},\ \bibinfo
  {pages} {012130} (\bibinfo {year} {2015})}\BibitemShut {NoStop}%
\bibitem [{\citenamefont {Polettini}\ and\ \citenamefont
  {Esposito}(2017)}]{polettini2017effective}%
  \BibitemOpen
  \bibfield  {author} {\bibinfo {author} {\bibfnamefont {M.}~\bibnamefont
  {Polettini}}\ and\ \bibinfo {author} {\bibfnamefont {M.}~\bibnamefont
  {Esposito}},\ }\bibfield  {title} {\bibinfo {title} {Effective thermodynamics
  for a marginal observer},\ }\href
  {https://doi.org/10.1103/PhysRevLett.119.240601} {\bibfield  {journal}
  {\bibinfo  {journal} {Phys. Rev. Lett.}\ }\textbf {\bibinfo {volume} {119}},\
  \bibinfo {pages} {240601} (\bibinfo {year} {2017})}\BibitemShut {NoStop}%
\bibitem [{\citenamefont {Bisker}\ \emph {et~al.}(2017)\citenamefont {Bisker},
  \citenamefont {Polettini}, \citenamefont {Gingrich},\ and\ \citenamefont
  {Horowitz}}]{bisker2017hierarchical}%
  \BibitemOpen
  \bibfield  {author} {\bibinfo {author} {\bibfnamefont {G.}~\bibnamefont
  {Bisker}}, \bibinfo {author} {\bibfnamefont {M.}~\bibnamefont {Polettini}},
  \bibinfo {author} {\bibfnamefont {T.~R.}\ \bibnamefont {Gingrich}},\ and\
  \bibinfo {author} {\bibfnamefont {J.~M.}\ \bibnamefont {Horowitz}},\
  }\bibfield  {title} {\bibinfo {title} {Hierarchical bounds on entropy
  production inferred from partial information},\ }\href
  {https://doi.org/10.1088/1742-5468/aa8c0d} {\bibfield  {journal} {\bibinfo
  {journal} {J. Stat. Mech.: Theory Exp.}\ }\textbf {\bibinfo {volume}
  {2017}}\bibinfo  {number} { (9)},\ \bibinfo {pages} {093210}}\BibitemShut
  {NoStop}%
\bibitem [{\citenamefont {Gingrich}\ \emph {et~al.}(2017)\citenamefont
  {Gingrich}, \citenamefont {Rotskoff},\ and\ \citenamefont
  {Horowitz}}]{gingrich2017inferring}%
  \BibitemOpen
\bibfield  {number} {  }\bibfield  {author} {\bibinfo {author} {\bibfnamefont
  {T.~R.}\ \bibnamefont {Gingrich}}, \bibinfo {author} {\bibfnamefont {G.~M.}\
  \bibnamefont {Rotskoff}},\ and\ \bibinfo {author} {\bibfnamefont {J.~M.}\
  \bibnamefont {Horowitz}},\ }\bibfield  {title} {\bibinfo {title} {Inferring
  dissipation from current fluctuations},\ }\href
  {https://doi.org/10.1088/1751-8121/aa672f} {\bibfield  {journal} {\bibinfo
  {journal} {J. Phys. A: Math. Theor.}\ }\textbf {\bibinfo {volume} {50}},\
  \bibinfo {pages} {184004} (\bibinfo {year} {2017})}\BibitemShut {NoStop}%
\bibitem [{\citenamefont {Dechant}\ and\ \citenamefont
  {Sasa}(2021)}]{dechant2021improving}%
  \BibitemOpen
  \bibfield  {author} {\bibinfo {author} {\bibfnamefont {A.}~\bibnamefont
  {Dechant}}\ and\ \bibinfo {author} {\bibfnamefont {S.-i.}\ \bibnamefont
  {Sasa}},\ }\bibfield  {title} {\bibinfo {title} {Improving thermodynamic
  bounds using correlations},\ }\href
  {https://doi.org/10.1103/PhysRevX.11.041061} {\bibfield  {journal} {\bibinfo
  {journal} {Phys. Rev. X}\ }\textbf {\bibinfo {volume} {11}},\ \bibinfo
  {pages} {041061} (\bibinfo {year} {2021})}\BibitemShut {NoStop}%
\bibitem [{\citenamefont {Lander}\ \emph {et~al.}(2012)\citenamefont {Lander},
  \citenamefont {Mehl}, \citenamefont {Blickle}, \citenamefont {Bechinger},\
  and\ \citenamefont {Seifert}}]{lander2012noninvasive}%
  \BibitemOpen
  \bibfield  {author} {\bibinfo {author} {\bibfnamefont {B.}~\bibnamefont
  {Lander}}, \bibinfo {author} {\bibfnamefont {J.}~\bibnamefont {Mehl}},
  \bibinfo {author} {\bibfnamefont {V.}~\bibnamefont {Blickle}}, \bibinfo
  {author} {\bibfnamefont {C.}~\bibnamefont {Bechinger}},\ and\ \bibinfo
  {author} {\bibfnamefont {U.}~\bibnamefont {Seifert}},\ }\bibfield  {title}
  {\bibinfo {title} {Noninvasive measurement of dissipation in colloidal
  systems},\ }\href {https://doi.org/10.1103/PhysRevE.86.030401} {\bibfield
  {journal} {\bibinfo  {journal} {Phys. Rev. E}\ }\textbf {\bibinfo {volume}
  {86}},\ \bibinfo {pages} {030401(R)} (\bibinfo {year} {2012})}\BibitemShut
  {NoStop}%
\bibitem [{\citenamefont {Frishman}\ and\ \citenamefont
  {Ronceray}(2020)}]{frishman2020learning}%
  \BibitemOpen
  \bibfield  {author} {\bibinfo {author} {\bibfnamefont {A.}~\bibnamefont
  {Frishman}}\ and\ \bibinfo {author} {\bibfnamefont {P.}~\bibnamefont
  {Ronceray}},\ }\bibfield  {title} {\bibinfo {title} {Learning force fields
  from stochastic trajectories},\ }\href
  {https://doi.org/10.1103/PhysRevX.10.021009} {\bibfield  {journal} {\bibinfo
  {journal} {Phys. Rev. X}\ }\textbf {\bibinfo {volume} {10}},\ \bibinfo
  {pages} {021009} (\bibinfo {year} {2020})}\BibitemShut {NoStop}%
\bibitem [{\citenamefont {Gnesotto}\ \emph {et~al.}(2020)\citenamefont
  {Gnesotto}, \citenamefont {Gradziuk}, \citenamefont {Ronceray},\ and\
  \citenamefont {Broedersz}}]{Gnesotto2020Learning}%
  \BibitemOpen
  \bibfield  {author} {\bibinfo {author} {\bibfnamefont {F.~S.}\ \bibnamefont
  {Gnesotto}}, \bibinfo {author} {\bibfnamefont {G.}~\bibnamefont {Gradziuk}},
  \bibinfo {author} {\bibfnamefont {P.}~\bibnamefont {Ronceray}},\ and\
  \bibinfo {author} {\bibfnamefont {C.~P.}\ \bibnamefont {Broedersz}},\
  }\bibfield  {title} {\bibinfo {title} {Learning the non-equilibrium dynamics
  of {B}rownian movies},\ }\href {https://doi.org/10.1038/s41467-020-18796-9}
  {\bibfield  {journal} {\bibinfo  {journal} {Nat. Commun.}\ }\textbf {\bibinfo
  {volume} {11}},\ \bibinfo {pages} {5378} (\bibinfo {year}
  {2020})}\BibitemShut {NoStop}%
\bibitem [{\citenamefont {Li}\ \emph {et~al.}(2019)\citenamefont {Li},
  \citenamefont {Horowitz}, \citenamefont {Gingrich},\ and\ \citenamefont
  {Fakhri}}]{Li2019Quantifying}%
  \BibitemOpen
  \bibfield  {author} {\bibinfo {author} {\bibfnamefont {J.}~\bibnamefont
  {Li}}, \bibinfo {author} {\bibfnamefont {J.~M.}\ \bibnamefont {Horowitz}},
  \bibinfo {author} {\bibfnamefont {T.~R.}\ \bibnamefont {Gingrich}},\ and\
  \bibinfo {author} {\bibfnamefont {N.}~\bibnamefont {Fakhri}},\ }\bibfield
  {title} {\bibinfo {title} {Quantifying dissipation using fluctuating
  currents},\ }\href {https://doi.org/10.1038/s41467-019-09631-x} {\bibfield
  {journal} {\bibinfo  {journal} {Nat. Commun.}\ }\textbf {\bibinfo {volume}
  {10}},\ \bibinfo {pages} {1666} (\bibinfo {year} {2019})}\BibitemShut
  {NoStop}%
\bibitem [{\citenamefont {Manikandan}\ \emph {et~al.}(2020)\citenamefont
  {Manikandan}, \citenamefont {Gupta},\ and\ \citenamefont
  {Krishnamurthy}}]{manikandan2020inferring}%
  \BibitemOpen
  \bibfield  {author} {\bibinfo {author} {\bibfnamefont {S.~K.}\ \bibnamefont
  {Manikandan}}, \bibinfo {author} {\bibfnamefont {D.}~\bibnamefont {Gupta}},\
  and\ \bibinfo {author} {\bibfnamefont {S.}~\bibnamefont {Krishnamurthy}},\
  }\bibfield  {title} {\bibinfo {title} {Inferring entropy production from
  short experiments},\ }\href {https://doi.org/10.1103/PhysRevLett.124.120603}
  {\bibfield  {journal} {\bibinfo  {journal} {Phys. Rev. Lett.}\ }\textbf
  {\bibinfo {volume} {124}},\ \bibinfo {pages} {120603} (\bibinfo {year}
  {2020})}\BibitemShut {NoStop}%
\bibitem [{\citenamefont {Otsubo}\ \emph {et~al.}(2020)\citenamefont {Otsubo},
  \citenamefont {Ito}, \citenamefont {Dechant},\ and\ \citenamefont
  {Sagawa}}]{otsubo2020estimating}%
  \BibitemOpen
  \bibfield  {author} {\bibinfo {author} {\bibfnamefont {S.}~\bibnamefont
  {Otsubo}}, \bibinfo {author} {\bibfnamefont {S.}~\bibnamefont {Ito}},
  \bibinfo {author} {\bibfnamefont {A.}~\bibnamefont {Dechant}},\ and\ \bibinfo
  {author} {\bibfnamefont {T.}~\bibnamefont {Sagawa}},\ }\bibfield  {title}
  {\bibinfo {title} {Estimating entropy production by machine learning of
  short-time fluctuating currents},\ }\href
  {https://doi.org/10.1103/PhysRevE.101.062106} {\bibfield  {journal} {\bibinfo
   {journal} {Phys. Rev. E}\ }\textbf {\bibinfo {volume} {101}},\ \bibinfo
  {pages} {062106} (\bibinfo {year} {2020})}\BibitemShut {NoStop}%
\bibitem [{\citenamefont {Van~Vu}\ \emph {et~al.}(2020)\citenamefont {Van~Vu},
  \citenamefont {Vo},\ and\ \citenamefont {Hasegawa}}]{vu2020entropy}%
  \BibitemOpen
  \bibfield  {author} {\bibinfo {author} {\bibfnamefont {T.}~\bibnamefont
  {Van~Vu}}, \bibinfo {author} {\bibfnamefont {V.~T.}\ \bibnamefont {Vo}},\
  and\ \bibinfo {author} {\bibfnamefont {Y.}~\bibnamefont {Hasegawa}},\
  }\bibfield  {title} {\bibinfo {title} {Entropy production estimation with
  optimal current},\ }\href {https://doi.org/10.1103/PhysRevE.101.042138}
  {\bibfield  {journal} {\bibinfo  {journal} {Phys. Rev. E}\ }\textbf {\bibinfo
  {volume} {101}},\ \bibinfo {pages} {042138} (\bibinfo {year}
  {2020})}\BibitemShut {NoStop}%
\bibitem [{\citenamefont {Manikandan}\ \emph {et~al.}(2021)\citenamefont
  {Manikandan}, \citenamefont {Ghosh}, \citenamefont {Kundu}, \citenamefont
  {Das}, \citenamefont {Agrawal}, \citenamefont {Mitra}, \citenamefont
  {Banerjee},\ and\ \citenamefont
  {Krishnamurthy}}]{Manikandan2021Quantitative}%
  \BibitemOpen
  \bibfield  {author} {\bibinfo {author} {\bibfnamefont {S.~K.}\ \bibnamefont
  {Manikandan}}, \bibinfo {author} {\bibfnamefont {S.}~\bibnamefont {Ghosh}},
  \bibinfo {author} {\bibfnamefont {A.}~\bibnamefont {Kundu}}, \bibinfo
  {author} {\bibfnamefont {B.}~\bibnamefont {Das}}, \bibinfo {author}
  {\bibfnamefont {V.}~\bibnamefont {Agrawal}}, \bibinfo {author} {\bibfnamefont
  {D.}~\bibnamefont {Mitra}}, \bibinfo {author} {\bibfnamefont
  {A.}~\bibnamefont {Banerjee}},\ and\ \bibinfo {author} {\bibfnamefont
  {S.}~\bibnamefont {Krishnamurthy}},\ }\bibfield  {title} {\bibinfo {title}
  {Quantitative analysis of non-equilibrium systems from short-time
  experimental data},\ }\href {https://doi.org/10.1038/s42005-021-00766-2}
  {\bibfield  {journal} {\bibinfo  {journal} {Commun. Phys.}\ }\textbf
  {\bibinfo {volume} {4}},\ \bibinfo {pages} {258} (\bibinfo {year}
  {2021})}\BibitemShut {NoStop}%
\bibitem [{\citenamefont {Kim}\ \emph {et~al.}(2020)\citenamefont {Kim},
  \citenamefont {Bae}, \citenamefont {Lee},\ and\ \citenamefont
  {Jeong}}]{Kim2020Learning}%
  \BibitemOpen
  \bibfield  {author} {\bibinfo {author} {\bibfnamefont {D.-K.}\ \bibnamefont
  {Kim}}, \bibinfo {author} {\bibfnamefont {Y.}~\bibnamefont {Bae}}, \bibinfo
  {author} {\bibfnamefont {S.}~\bibnamefont {Lee}},\ and\ \bibinfo {author}
  {\bibfnamefont {H.}~\bibnamefont {Jeong}},\ }\bibfield  {title} {\bibinfo
  {title} {Learning entropy production via neural networks},\ }\href
  {https://doi.org/10.1103/PhysRevLett.125.140604} {\bibfield  {journal}
  {\bibinfo  {journal} {Phys. Rev. Lett.}\ }\textbf {\bibinfo {volume} {125}},\
  \bibinfo {pages} {140604} (\bibinfo {year} {2020})}\BibitemShut {NoStop}%
\bibitem [{\citenamefont {Kim}\ \emph {et~al.}(2022)\citenamefont {Kim},
  \citenamefont {Lee},\ and\ \citenamefont {Jeong}}]{Kim2022Estimating}%
  \BibitemOpen
  \bibfield  {author} {\bibinfo {author} {\bibfnamefont {D.-K.}\ \bibnamefont
  {Kim}}, \bibinfo {author} {\bibfnamefont {S.}~\bibnamefont {Lee}},\ and\
  \bibinfo {author} {\bibfnamefont {H.}~\bibnamefont {Jeong}},\ }\bibfield
  {title} {\bibinfo {title} {Estimating entropy production with odd-parity
  state variables via machine learning},\ }\href
  {https://doi.org/10.1103/PhysRevResearch.4.023051} {\bibfield  {journal}
  {\bibinfo  {journal} {Phys. Rev. Res.}\ }\textbf {\bibinfo {volume} {4}},\
  \bibinfo {pages} {023051} (\bibinfo {year} {2022})}\BibitemShut {NoStop}%
\bibitem [{\citenamefont {Otsubo}\ \emph {et~al.}(2022)\citenamefont {Otsubo},
  \citenamefont {Manikandan}, \citenamefont {Sagawa},\ and\ \citenamefont
  {Krishnamurthy}}]{Otsubo2022estimating}%
  \BibitemOpen
  \bibfield  {author} {\bibinfo {author} {\bibfnamefont {S.}~\bibnamefont
  {Otsubo}}, \bibinfo {author} {\bibfnamefont {S.~K.}\ \bibnamefont
  {Manikandan}}, \bibinfo {author} {\bibfnamefont {T.}~\bibnamefont {Sagawa}},\
  and\ \bibinfo {author} {\bibfnamefont {S.}~\bibnamefont {Krishnamurthy}},\
  }\bibfield  {title} {\bibinfo {title} {Estimating time-dependent entropy
  production from non-equilibrium trajectories},\ }\href
  {https://doi.org/10.1038/s42005-021-00787-x} {\bibfield  {journal} {\bibinfo
  {journal} {Commun. Phys.}\ }\textbf {\bibinfo {volume} {5}},\ \bibinfo
  {pages} {11} (\bibinfo {year} {2022})}\BibitemShut {NoStop}%
\bibitem [{\citenamefont {Lee}\ \emph {et~al.}(2023)\citenamefont {Lee},
  \citenamefont {Kim}, \citenamefont {Park}, \citenamefont {Kim}, \citenamefont
  {Park},\ and\ \citenamefont {Lee}}]{Lee2023Multidimensional}%
  \BibitemOpen
  \bibfield  {author} {\bibinfo {author} {\bibfnamefont {S.}~\bibnamefont
  {Lee}}, \bibinfo {author} {\bibfnamefont {D.-K.}\ \bibnamefont {Kim}},
  \bibinfo {author} {\bibfnamefont {J.-M.}\ \bibnamefont {Park}}, \bibinfo
  {author} {\bibfnamefont {W.~K.}\ \bibnamefont {Kim}}, \bibinfo {author}
  {\bibfnamefont {H.}~\bibnamefont {Park}},\ and\ \bibinfo {author}
  {\bibfnamefont {J.~S.}\ \bibnamefont {Lee}},\ }\bibfield  {title} {\bibinfo
  {title} {Multidimensional entropic bound: Estimator of entropy production for
  {L}angevin dynamics with an arbitrary time-dependent protocol},\ }\href
  {https://doi.org/10.1103/PhysRevResearch.5.013194} {\bibfield  {journal}
  {\bibinfo  {journal} {Phys. Rev. Res.}\ }\textbf {\bibinfo {volume} {5}},\
  \bibinfo {pages} {013194} (\bibinfo {year} {2023})}\BibitemShut {NoStop}%
\bibitem [{\citenamefont {Kappler}\ and\ \citenamefont
  {Adhikari}(2022)}]{Kappler2022Measurement}%
  \BibitemOpen
  \bibfield  {author} {\bibinfo {author} {\bibfnamefont {J.}~\bibnamefont
  {Kappler}}\ and\ \bibinfo {author} {\bibfnamefont {R.}~\bibnamefont
  {Adhikari}},\ }\bibfield  {title} {\bibinfo {title} {Measurement of
  irreversibility and entropy production via the tubular ensemble},\ }\href
  {https://doi.org/10.1103/PhysRevE.105.044107} {\bibfield  {journal} {\bibinfo
   {journal} {Phys. Rev. E}\ }\textbf {\bibinfo {volume} {105}},\ \bibinfo
  {pages} {044107} (\bibinfo {year} {2022})}\BibitemShut {NoStop}%
\bibitem [{\citenamefont {Ito}\ and\ \citenamefont
  {Dechant}(2020)}]{ito2020prx}%
  \BibitemOpen
  \bibfield  {author} {\bibinfo {author} {\bibfnamefont {S.}~\bibnamefont
  {Ito}}\ and\ \bibinfo {author} {\bibfnamefont {A.}~\bibnamefont {Dechant}},\
  }\bibfield  {title} {\bibinfo {title} {Stochastic time evolution, information
  geometry, and the {C}ram\'er-{R}ao bound},\ }\href
  {https://doi.org/10.1103/PhysRevX.10.021056} {\bibfield  {journal} {\bibinfo
  {journal} {Phys. Rev. X}\ }\textbf {\bibinfo {volume} {10}},\ \bibinfo
  {pages} {021056} (\bibinfo {year} {2020})}\BibitemShut {NoStop}%
\bibitem [{\citenamefont {Mart{\'\i}nez}\ \emph {et~al.}(2019)\citenamefont
  {Mart{\'\i}nez}, \citenamefont {Bisker}, \citenamefont {Horowitz},\ and\
  \citenamefont {Parrondo}}]{martinez2019inferring}%
  \BibitemOpen
  \bibfield  {author} {\bibinfo {author} {\bibfnamefont {I.~A.}\ \bibnamefont
  {Mart{\'\i}nez}}, \bibinfo {author} {\bibfnamefont {G.}~\bibnamefont
  {Bisker}}, \bibinfo {author} {\bibfnamefont {J.~M.}\ \bibnamefont
  {Horowitz}},\ and\ \bibinfo {author} {\bibfnamefont {J.~M.}\ \bibnamefont
  {Parrondo}},\ }\bibfield  {title} {\bibinfo {title} {Inferring broken
  detailed balance in the absence of observable currents},\ }\href
  {https://doi.org/10.1038/s41467-019-11051-w} {\bibfield  {journal} {\bibinfo
  {journal} {Nat. Commun.}\ }\textbf {\bibinfo {volume} {10}},\ \bibinfo
  {pages} {3542} (\bibinfo {year} {2019})}\BibitemShut {NoStop}%
\bibitem [{\citenamefont {Skinner}\ and\ \citenamefont
  {Dunkel}(2021{\natexlab{b}})}]{skinner2021estimating}%
  \BibitemOpen
  \bibfield  {author} {\bibinfo {author} {\bibfnamefont {D.~J.}\ \bibnamefont
  {Skinner}}\ and\ \bibinfo {author} {\bibfnamefont {J.}~\bibnamefont
  {Dunkel}},\ }\bibfield  {title} {\bibinfo {title} {Estimating entropy
  production from waiting time distributions},\ }\href
  {https://doi.org/10.1103/PhysRevLett.127.198101} {\bibfield  {journal}
  {\bibinfo  {journal} {Phys. Rev. Lett.}\ }\textbf {\bibinfo {volume} {127}},\
  \bibinfo {pages} {198101} (\bibinfo {year} {2021}{\natexlab{b}})}\BibitemShut
  {NoStop}%
\bibitem [{\citenamefont {van~der Meer}\ \emph {et~al.}(2022)\citenamefont
  {van~der Meer}, \citenamefont {Ertel},\ and\ \citenamefont
  {Seifert}}]{vandermeer2022thermodynamic}%
  \BibitemOpen
  \bibfield  {author} {\bibinfo {author} {\bibfnamefont {J.}~\bibnamefont
  {van~der Meer}}, \bibinfo {author} {\bibfnamefont {B.}~\bibnamefont
  {Ertel}},\ and\ \bibinfo {author} {\bibfnamefont {U.}~\bibnamefont
  {Seifert}},\ }\bibfield  {title} {\bibinfo {title} {Thermodynamic inference
  in partially accessible {M}arkov networks: A unifying perspective from
  transition-based waiting time distributions},\ }\href
  {https://doi.org/10.1103/PhysRevX.12.031025} {\bibfield  {journal} {\bibinfo
  {journal} {Phys. Rev. X}\ }\textbf {\bibinfo {volume} {12}},\ \bibinfo
  {pages} {031025} (\bibinfo {year} {2022})}\BibitemShut {NoStop}%
\bibitem [{\citenamefont {Harunari}\ \emph {et~al.}(2022)\citenamefont
  {Harunari}, \citenamefont {Dutta}, \citenamefont {Polettini},\ and\
  \citenamefont {Rold\'an}}]{harunari2022learn}%
  \BibitemOpen
  \bibfield  {author} {\bibinfo {author} {\bibfnamefont {P.~E.}\ \bibnamefont
  {Harunari}}, \bibinfo {author} {\bibfnamefont {A.}~\bibnamefont {Dutta}},
  \bibinfo {author} {\bibfnamefont {M.}~\bibnamefont {Polettini}},\ and\
  \bibinfo {author} {\bibfnamefont {E.}~\bibnamefont {Rold\'an}},\ }\bibfield
  {title} {\bibinfo {title} {What to learn from a few visible transitions'
  statistics?},\ }\href {https://doi.org/10.1103/PhysRevX.12.041026} {\bibfield
   {journal} {\bibinfo  {journal} {Phys. Rev. X}\ }\textbf {\bibinfo {volume}
  {12}},\ \bibinfo {pages} {041026} (\bibinfo {year} {2022})}\BibitemShut
  {NoStop}%
\bibitem [{\citenamefont {Amari}\ and\ \citenamefont
  {Nagaoka}(2000)}]{amari2007methods}%
  \BibitemOpen
  \bibfield  {author} {\bibinfo {author} {\bibfnamefont {S.-i.}\ \bibnamefont
  {Amari}}\ and\ \bibinfo {author} {\bibfnamefont {H.}~\bibnamefont
  {Nagaoka}},\ }\href@noop {} {\emph {\bibinfo {title} {{Methods of Information
  Geometry}}}}\ (\bibinfo  {publisher} {American Mathematical Society/Oxford
  University Press},\ \bibinfo {address} {Providence},\ \bibinfo {year}
  {2000})\ \bibinfo {note} {(Originally published in Japanese, Iwanami, Tokyo,
  1993)}\BibitemShut {NoStop}%
\bibitem [{\citenamefont {Amari}(2016)}]{amari2016information}%
  \BibitemOpen
  \bibfield  {author} {\bibinfo {author} {\bibfnamefont {S.-i.}\ \bibnamefont
  {Amari}},\ }\href@noop {} {\emph {\bibinfo {title} {{Information Geometry and
  Its Applications}}}}\ (\bibinfo  {publisher} {Springer},\ \bibinfo {address}
  {Tokyo},\ \bibinfo {year} {2016})\BibitemShut {NoStop}%
\bibitem [{\citenamefont {{Van den Broeck}}\ and\ \citenamefont
  {Esposito}(2015)}]{vandenbroeck2015ensemble}%
  \BibitemOpen
  \bibfield  {author} {\bibinfo {author} {\bibfnamefont {C.}~\bibnamefont {{Van
  den Broeck}}}\ and\ \bibinfo {author} {\bibfnamefont {M.}~\bibnamefont
  {Esposito}},\ }\bibfield  {title} {\bibinfo {title} {Ensemble and trajectory
  thermodynamics: A brief introduction},\ }\href
  {https://doi.org/https://doi.org/10.1016/j.physa.2014.04.035} {\bibfield
  {journal} {\bibinfo  {journal} {Physica (Amsterdam)}\ }\textbf {\bibinfo
  {volume} {418A}},\ \bibinfo {pages} {6} (\bibinfo {year} {2015})}\BibitemShut
  {NoStop}%
\bibitem [{\citenamefont {Strasberg}\ and\ \citenamefont
  {Esposito}(2019)}]{Strasberg2019NonMarkovianity}%
  \BibitemOpen
  \bibfield  {author} {\bibinfo {author} {\bibfnamefont {P.}~\bibnamefont
  {Strasberg}}\ and\ \bibinfo {author} {\bibfnamefont {M.}~\bibnamefont
  {Esposito}},\ }\bibfield  {title} {\bibinfo {title} {Non-{M}arkovianity and
  negative entropy production rates},\ }\href
  {https://doi.org/10.1103/PhysRevE.99.012120} {\bibfield  {journal} {\bibinfo
  {journal} {Phys. Rev. E}\ }\textbf {\bibinfo {volume} {99}},\ \bibinfo
  {pages} {012120} (\bibinfo {year} {2019})}\BibitemShut {NoStop}%
\bibitem [{\citenamefont {Mehta}\ \emph {et~al.}(1999)\citenamefont {Mehta},
  \citenamefont {Rief}, \citenamefont {Spudich}, \citenamefont {Smith},\ and\
  \citenamefont {Simmons}}]{mehta1999single}%
  \BibitemOpen
  \bibfield  {author} {\bibinfo {author} {\bibfnamefont {A.~D.}\ \bibnamefont
  {Mehta}}, \bibinfo {author} {\bibfnamefont {M.}~\bibnamefont {Rief}},
  \bibinfo {author} {\bibfnamefont {J.~A.}\ \bibnamefont {Spudich}}, \bibinfo
  {author} {\bibfnamefont {D.~A.}\ \bibnamefont {Smith}},\ and\ \bibinfo
  {author} {\bibfnamefont {R.~M.}\ \bibnamefont {Simmons}},\ }\bibfield
  {title} {\bibinfo {title} {Single-molecule biomechanics with optical
  methods},\ }\href {https://doi.org/10.1126/science.283.5408.1689} {\bibfield
  {journal} {\bibinfo  {journal} {Science}\ }\textbf {\bibinfo {volume}
  {283}},\ \bibinfo {pages} {1689} (\bibinfo {year} {1999})}\BibitemShut
  {NoStop}%
\bibitem [{\citenamefont {van Kampen}(2007)}]{kampen2007stochastic}%
  \BibitemOpen
  \bibfield  {author} {\bibinfo {author} {\bibfnamefont {N.~G.}\ \bibnamefont
  {van Kampen}},\ }\href@noop {} {\emph {\bibinfo {title} {Stochastic Processes
  in Physics and Chemistry}}},\ \bibinfo {edition} {3rd}\ ed.\ (\bibinfo
  {publisher} {North-Holland},\ \bibinfo {address} {Amsterdam},\ \bibinfo
  {year} {2007})\BibitemShut {NoStop}%
\bibitem [{\citenamefont {Van~den Broeck}\ and\ \citenamefont
  {Esposito}(2010)}]{vandenbroeck2010three}%
  \BibitemOpen
  \bibfield  {author} {\bibinfo {author} {\bibfnamefont {C.}~\bibnamefont
  {Van~den Broeck}}\ and\ \bibinfo {author} {\bibfnamefont {M.}~\bibnamefont
  {Esposito}},\ }\bibfield  {title} {\bibinfo {title} {Three faces of the
  second law. {II}. {F}okker-{P}lanck formulation},\ }\href
  {https://doi.org/10.1103/PhysRevE.82.011144} {\bibfield  {journal} {\bibinfo
  {journal} {Phys. Rev. E}\ }\textbf {\bibinfo {volume} {82}},\ \bibinfo
  {pages} {011144} (\bibinfo {year} {2010})}\BibitemShut {NoStop}%
\bibitem [{\citenamefont {Kullback}\ and\ \citenamefont
  {Leibler}(1951)}]{kullback1951information}%
  \BibitemOpen
  \bibfield  {author} {\bibinfo {author} {\bibfnamefont {S.}~\bibnamefont
  {Kullback}}\ and\ \bibinfo {author} {\bibfnamefont {R.~A.}\ \bibnamefont
  {Leibler}},\ }\bibfield  {title} {\bibinfo {title} {On information and
  sufficiency},\ }\href {https://doi.org/10.1214/aoms/1177729694} {\bibfield
  {journal} {\bibinfo  {journal} {Ann. Math. Stat.}\ }\textbf {\bibinfo
  {volume} {22}},\ \bibinfo {pages} {79} (\bibinfo {year} {1951})}\BibitemShut
  {NoStop}%
\bibitem [{\citenamefont {Dechant}\ and\ \citenamefont
  {Sasa}(2018)}]{Dechant2018CurrentFluctuations}%
  \BibitemOpen
  \bibfield  {author} {\bibinfo {author} {\bibfnamefont {A.}~\bibnamefont
  {Dechant}}\ and\ \bibinfo {author} {\bibfnamefont {S.-i.}\ \bibnamefont
  {Sasa}},\ }\bibfield  {title} {\bibinfo {title} {Current fluctuations and
  transport efficiency for general {L}angevin systems},\ }\href
  {https://doi.org/10.1088/1742-5468/aac91a} {\bibfield  {journal} {\bibinfo
  {journal} {J. Stat. Mech.: Theory Exp.}\ }\textbf {\bibinfo {volume}
  {2018}}\bibinfo  {number} { (6)},\ \bibinfo {pages} {063209}}\BibitemShut
  {NoStop}%
\bibitem [{\citenamefont {Liu}\ \emph {et~al.}(2020)\citenamefont {Liu},
  \citenamefont {Gong},\ and\ \citenamefont
  {Ueda}}]{Liu2020ThermodynamicUncertainty}%
  \BibitemOpen
\bibfield  {number} {  }\bibfield  {author} {\bibinfo {author} {\bibfnamefont
  {K.}~\bibnamefont {Liu}}, \bibinfo {author} {\bibfnamefont {Z.}~\bibnamefont
  {Gong}},\ and\ \bibinfo {author} {\bibfnamefont {M.}~\bibnamefont {Ueda}},\
  }\bibfield  {title} {\bibinfo {title} {Thermodynamic uncertainty relation for
  arbitrary initial states},\ }\href
  {https://doi.org/10.1103/PhysRevLett.125.140602} {\bibfield  {journal}
  {\bibinfo  {journal} {Phys. Rev. Lett.}\ }\textbf {\bibinfo {volume} {125}},\
  \bibinfo {pages} {140602} (\bibinfo {year} {2020})}\BibitemShut {NoStop}%
\bibitem [{\citenamefont {Koyuk}\ and\ \citenamefont
  {Seifert}(2020)}]{Koyuk2020Thermodynamic}%
  \BibitemOpen
  \bibfield  {author} {\bibinfo {author} {\bibfnamefont {T.}~\bibnamefont
  {Koyuk}}\ and\ \bibinfo {author} {\bibfnamefont {U.}~\bibnamefont
  {Seifert}},\ }\bibfield  {title} {\bibinfo {title} {Thermodynamic uncertainty
  relation for time-dependent driving},\ }\href
  {https://doi.org/10.1103/PhysRevLett.125.260604} {\bibfield  {journal}
  {\bibinfo  {journal} {Phys. Rev. Lett.}\ }\textbf {\bibinfo {volume} {125}},\
  \bibinfo {pages} {260604} (\bibinfo {year} {2020})}\BibitemShut {NoStop}%
\bibitem [{\citenamefont {Commons}\ \emph {et~al.}()\citenamefont {Commons},
  \citenamefont {Yang},\ and\ \citenamefont {Qian}}]{commons2021duality}%
  \BibitemOpen
  \bibfield  {author} {\bibinfo {author} {\bibfnamefont {J.}~\bibnamefont
  {Commons}}, \bibinfo {author} {\bibfnamefont {Y.-J.}\ \bibnamefont {Yang}},\
  and\ \bibinfo {author} {\bibfnamefont {H.}~\bibnamefont {Qian}},\ }\bibfield
  {title} {\bibinfo {title} {Duality symmetry, two entropy functions, and an
  eigenvalue problem in {G}ibbs' theory},\ }\href@noop {} {\bibinfo  {journal}
  {\href{https://arxiv.org/abs/2108.08948}{arXiv preprint arXiv:2108.08948
  (2021)}}\ }\BibitemShut {NoStop}%
\bibitem [{\citenamefont {Weinhold}(1975)}]{weinhold1975metric}%
  \BibitemOpen
\bibfield  {journal} {  }\bibfield  {author} {\bibinfo {author} {\bibfnamefont
  {F.}~\bibnamefont {Weinhold}},\ }\bibfield  {title} {\bibinfo {title} {Metric
  geometry of equilibrium thermodynamics},\ }\href
  {https://doi.org/10.1063/1.431689} {\bibfield  {journal} {\bibinfo  {journal}
  {J. Chem. Phys.}\ }\textbf {\bibinfo {volume} {63}},\ \bibinfo {pages} {2479}
  (\bibinfo {year} {1975})}\BibitemShut {NoStop}%
\bibitem [{\citenamefont {Ruppeiner}(1979)}]{ruppeiner1979thermodynamics}%
  \BibitemOpen
  \bibfield  {author} {\bibinfo {author} {\bibfnamefont {G.}~\bibnamefont
  {Ruppeiner}},\ }\bibfield  {title} {\bibinfo {title} {Thermodynamics: {A}
  {R}iemannian geometric model},\ }\href
  {https://doi.org/10.1103/PhysRevA.20.1608} {\bibfield  {journal} {\bibinfo
  {journal} {Phys. Rev. A}\ }\textbf {\bibinfo {volume} {20}},\ \bibinfo
  {pages} {1608} (\bibinfo {year} {1979})}\BibitemShut {NoStop}%
\bibitem [{\citenamefont {Crooks}(2007)}]{crooks2007measuring}%
  \BibitemOpen
  \bibfield  {author} {\bibinfo {author} {\bibfnamefont {G.~E.}\ \bibnamefont
  {Crooks}},\ }\bibfield  {title} {\bibinfo {title} {Measuring thermodynamic
  length},\ }\href {https://doi.org/10.1103/PhysRevLett.99.100602} {\bibfield
  {journal} {\bibinfo  {journal} {Phys. Rev. Lett.}\ }\textbf {\bibinfo
  {volume} {99}},\ \bibinfo {pages} {100602} (\bibinfo {year}
  {2007})}\BibitemShut {NoStop}%
\bibitem [{\citenamefont {Ito}(2018)}]{ito2018prl}%
  \BibitemOpen
  \bibfield  {author} {\bibinfo {author} {\bibfnamefont {S.}~\bibnamefont
  {Ito}},\ }\bibfield  {title} {\bibinfo {title} {Stochastic thermodynamic
  interpretation of information geometry},\ }\href
  {https://doi.org/10.1103/PhysRevLett.121.030605} {\bibfield  {journal}
  {\bibinfo  {journal} {Phys. Rev. Lett.}\ }\textbf {\bibinfo {volume} {121}},\
  \bibinfo {pages} {030605} (\bibinfo {year} {2018})}\BibitemShut {NoStop}%
\bibitem [{\citenamefont {Kolchinsky}\ and\ \citenamefont
  {Wolpert}(2021)}]{kolchinsky2021prx}%
  \BibitemOpen
  \bibfield  {author} {\bibinfo {author} {\bibfnamefont {A.}~\bibnamefont
  {Kolchinsky}}\ and\ \bibinfo {author} {\bibfnamefont {D.~H.}\ \bibnamefont
  {Wolpert}},\ }\bibfield  {title} {\bibinfo {title} {Work, entropy production,
  and thermodynamics of information under protocol constraints},\ }\href
  {https://doi.org/10.1103/PhysRevX.11.041024} {\bibfield  {journal} {\bibinfo
  {journal} {Phys. Rev. X}\ }\textbf {\bibinfo {volume} {11}},\ \bibinfo
  {pages} {041024} (\bibinfo {year} {2021})}\BibitemShut {NoStop}%
\bibitem [{\citenamefont {Vaikuntanathan}\ and\ \citenamefont
  {Jarzynski}(2009)}]{Vaikuntanathan2009DissipationLag}%
  \BibitemOpen
  \bibfield  {author} {\bibinfo {author} {\bibfnamefont {S.}~\bibnamefont
  {Vaikuntanathan}}\ and\ \bibinfo {author} {\bibfnamefont {C.}~\bibnamefont
  {Jarzynski}},\ }\bibfield  {title} {\bibinfo {title} {Dissipation and lag in
  irreversible processes},\ }\href {https://doi.org/10.1209/0295-5075/87/60005}
  {\bibfield  {journal} {\bibinfo  {journal} {EPL}\ }\textbf {\bibinfo {volume}
  {87}},\ \bibinfo {pages} {60005} (\bibinfo {year} {2009})}\BibitemShut
  {NoStop}%
\bibitem [{\citenamefont {Takara}\ \emph {et~al.}(2010)\citenamefont {Takara},
  \citenamefont {Hasegawa},\ and\ \citenamefont
  {Driebe}}]{takara2010generalization}%
  \BibitemOpen
  \bibfield  {author} {\bibinfo {author} {\bibfnamefont {K.}~\bibnamefont
  {Takara}}, \bibinfo {author} {\bibfnamefont {H.-H.}\ \bibnamefont
  {Hasegawa}},\ and\ \bibinfo {author} {\bibfnamefont {D.}~\bibnamefont
  {Driebe}},\ }\bibfield  {title} {\bibinfo {title} {Generalization of the
  second law for a transition between nonequilibrium states},\ }\href
  {https://doi.org/10.1016/j.physleta.2010.11.002} {\bibfield  {journal}
  {\bibinfo  {journal} {Phys. Lett. A}\ }\textbf {\bibinfo {volume} {375}},\
  \bibinfo {pages} {88} (\bibinfo {year} {2010})}\BibitemShut {NoStop}%
\bibitem [{\citenamefont {Esposito}\ and\ \citenamefont {Van~den
  Broeck}(2011)}]{esposito2011EPL}%
  \BibitemOpen
  \bibfield  {author} {\bibinfo {author} {\bibfnamefont {M.}~\bibnamefont
  {Esposito}}\ and\ \bibinfo {author} {\bibfnamefont {C.}~\bibnamefont {Van~den
  Broeck}},\ }\bibfield  {title} {\bibinfo {title} {Second law and {L}andauer
  principle far from equilibrium},\ }\href
  {https://doi.org/10.1209/0295-5075/95/40004} {\bibfield  {journal} {\bibinfo
  {journal} {EPL}\ }\textbf {\bibinfo {volume} {95}},\ \bibinfo {pages} {40004}
  (\bibinfo {year} {2011})}\BibitemShut {NoStop}%
\bibitem [{\citenamefont {Gorban}\ \emph {et~al.}(2010)\citenamefont {Gorban},
  \citenamefont {Gorban},\ and\ \citenamefont {Judge}}]{gorban2010entropy}%
  \BibitemOpen
  \bibfield  {author} {\bibinfo {author} {\bibfnamefont {A.~N.}\ \bibnamefont
  {Gorban}}, \bibinfo {author} {\bibfnamefont {P.~A.}\ \bibnamefont {Gorban}},\
  and\ \bibinfo {author} {\bibfnamefont {G.}~\bibnamefont {Judge}},\ }\bibfield
   {title} {\bibinfo {title} {Entropy: The {M}arkov ordering approach},\ }\href
  {https://doi.org/10.3390/e12051145} {\bibfield  {journal} {\bibinfo
  {journal} {Entropy}\ }\textbf {\bibinfo {volume} {12}},\ \bibinfo {pages}
  {1145} (\bibinfo {year} {2010})}\BibitemShut {NoStop}%
\bibitem [{\citenamefont {Donsker}\ and\ \citenamefont
  {Varadhan}(1983)}]{donsker1983asymptotic}%
  \BibitemOpen
  \bibfield  {author} {\bibinfo {author} {\bibfnamefont {M.~D.}\ \bibnamefont
  {Donsker}}\ and\ \bibinfo {author} {\bibfnamefont {S.~S.}\ \bibnamefont
  {Varadhan}},\ }\bibfield  {title} {\bibinfo {title} {Asymptotic evaluation of
  certain {M}arkov process expectations for large time. {IV}},\ }\href
  {https://doi.org/10.1002/cpa.3160360204} {\bibfield  {journal} {\bibinfo
  {journal} {Commun. Pure Appl. Math.}\ }\textbf {\bibinfo {volume} {36}},\
  \bibinfo {pages} {183} (\bibinfo {year} {1983})}\BibitemShut {NoStop}%
\bibitem [{\citenamefont {Ruelle}(1969)}]{ruelle1969}%
  \BibitemOpen
  \bibfield  {author} {\bibinfo {author} {\bibfnamefont {D.}~\bibnamefont
  {Ruelle}},\ }\href@noop {} {\emph {\bibinfo {title} {Statistical
  Mechanics}}}\ (\bibinfo  {publisher} {W. A. Benjamin},\ \bibinfo {address}
  {New York},\ \bibinfo {year} {1969})\BibitemShut {NoStop}%
\bibitem [{\citenamefont {Callen}(1985)}]{callen1985}%
  \BibitemOpen
  \bibfield  {author} {\bibinfo {author} {\bibfnamefont {H.~B.}\ \bibnamefont
  {Callen}},\ }\href@noop {} {\emph {\bibinfo {title} {Thermodynamics and an
  Introduction to Thermostatistics}}},\ \bibinfo {edition} {2nd}\ ed.\
  (\bibinfo  {publisher} {Wiley},\ \bibinfo {address} {New York},\ \bibinfo
  {year} {1985})\BibitemShut {NoStop}%
\bibitem [{\citenamefont {Hatano}\ and\ \citenamefont
  {Sasa}(2001)}]{hatano2001steady}%
  \BibitemOpen
  \bibfield  {author} {\bibinfo {author} {\bibfnamefont {T.}~\bibnamefont
  {Hatano}}\ and\ \bibinfo {author} {\bibfnamefont {S.-i.}\ \bibnamefont
  {Sasa}},\ }\bibfield  {title} {\bibinfo {title} {Steady-state thermodynamics
  of {L}angevin systems},\ }\href {https://doi.org/10.1103/PhysRevLett.86.3463}
  {\bibfield  {journal} {\bibinfo  {journal} {Phys. Rev. Lett.}\ }\textbf
  {\bibinfo {volume} {86}},\ \bibinfo {pages} {3463} (\bibinfo {year}
  {2001})}\BibitemShut {NoStop}%
\bibitem [{\citenamefont {Esposito}\ and\ \citenamefont {Van~den
  Broeck}(2010)}]{esposito2010three}%
  \BibitemOpen
  \bibfield  {author} {\bibinfo {author} {\bibfnamefont {M.}~\bibnamefont
  {Esposito}}\ and\ \bibinfo {author} {\bibfnamefont {C.}~\bibnamefont {Van~den
  Broeck}},\ }\bibfield  {title} {\bibinfo {title} {Three faces of the second
  law. {I}. {M}aster equation formulation},\ }\href
  {https://doi.org/10.1103/PhysRevE.82.011143} {\bibfield  {journal} {\bibinfo
  {journal} {Phys. Rev. E}\ }\textbf {\bibinfo {volume} {82}},\ \bibinfo
  {pages} {011143} (\bibinfo {year} {2010})}\BibitemShut {NoStop}%
\end{thebibliography}
\end{document}